\documentclass[aps,prc,twocolumn,showpacs]{revtex4}
\usepackage{amsmath}
\usepackage{ulem}
\usepackage{color}
\usepackage{graphicx}
\usepackage{epsfig}
\usepackage{subfig}
\usepackage{bm}
\def\be {\begin{equation}}
\def\ee {\end{equation}}
\def\nn {\nonumber}
\def\bea {\begin{eqnarray}}
\def\eea {\end{eqnarray}}

\newcommand{\ep}{\epsilon}
\newcommand{\om}{\omega}

\newcommand{\vp}{\vec p}
\newcommand{\del}{\partial}
\newcommand{\bp}{\boldsymbol{p}}
\begin{document}

%\linenumbers

\title{Anisotropic transport properties of Hadron Resonance Gas in magnetic field}% due to the magnetic field}

\author{Ashutosh Dash$^1$, Subhasis Samanta$^{2}$, Jayanta Dey$^3$, Utsab Gangopadhyaya$^1$, Sabyasachi Ghosh$^3$, Victor Roy$^1$}

% \email{jayantad@iitbhilai.ac.in}
% \email{subhasis.samant@gmail.com}
% \email{sabyaphy@gmail.com}
%
\affiliation{$^{1}$School of Physical Sciences, National Institute of Science 
Education and Research,Bhubaneswar, HBNI, Jatni, 752050, India}
\affiliation{$^2$Institute of Physics, Jan Kochanowski University, 25-406 Kielce, Poland}
\affiliation{$^{3}$Indian Institute of Technology Bhilai, GEC Campus, Sejbahar, Raipur 492015, Chhattisgarh, India}
%

%
% \author{Sabyasachi Ghosh}
% \email{sabyaphy@gmail.com}
% \affiliation{Indian Institute of Technology Bhilai, GEC Campus, Sejbahar, Raipur 492015, Chhattisgarh, India}

\begin{abstract}
An intense transient magnetic field is produced in high energy heavy-ion collisions
mostly due to the spectator protons inside the two colliding nucleus. 
The magnetic field introduces anisotropy in the medium and hence the isotropic scalar transport coefficients become anisotropic 
and split into multiple components. Here we calculate the anisotropic transport coefficients 
shear, bulk viscosity, electrical conductivity, and the thermal diffusion coefficients for a multicomponent 
Hadron-Resonance-Gas (HRG) model for a non-zero magnetic field by using the Boltzmann transport equation in a relaxation time approximation (RTA). 
The anisotropic transport coefficient component along the magnetic field remains unaffected by the magnetic field, while perpendicular 
dissipation is governed by the interplay of the collisional relaxation time and the magnetic time scale, which is inverse of the cyclotron frequency.
We calculate the anisotropic transport coefficients as a function of temperature and magnetic field using the HRG model.
The neutral hadrons are unaffected by the Lorentz force and do not contribute to the anisotropic transports, we estimate
within the HRG model the relative contribution of isotropic and anisotropic transports as a function of magnetic field and temperature. We also give an estimation of these anisotropic transport coefficients for the hadronic gas at finite baryon chemical potential ~($\mu_{B}$).

\end{abstract}

%\pacs{12.38.Mh,25.75.-q,24.85.+p,25.75.Nq}
\maketitle
\section{Introduction}
% Extremely strong magnetic fields have been known to exist during the 
% electroweak phase transition of the universe as suggested by cosmological models ~\cite{Vachaspati}
% Large values of magnetic fields are also present in the interior of dense neutron 
% stars called magnetars ~\cite{Duncan}. 
% Studying quantum field theory in the presence of magnetic field has led to many 
% interesting observations such as magnetic catalysis ~\cite{Shovkovy}, chiral
% magnetic effect ~\cite{Fukushima}, inverse magnetic catalysis ~\cite{Bali1,Bali2} and many more.
%

 In the initial stage of heavy ion collisions  an intense transient 
 magnetic field $eB\sim (1-10) m_{\pi}^2$(for \(\sqrt{s_{\mathrm{NN}}}=200\) GeV collisions)
is expected to be produced \cite{Bzdak:2011yy,Deng:2012pc,Tuchin:2013ie,Li:2016tel,Roy:2015coa}. 
Theoretically it was also shown that the magnitude of the 
magnetic field almost linearly rise with center of mass energy collisions \cite{Deng:2012pc,Tuchin:2013ie}. 

A general consensus is that the initial large magnetic field will decay quickly (within a few fm)
and becomes so weak that its effect may be negligible in any bulk observables. However,
the initial hot and dense phase of Quark-Gluon-Plasma~(QGP) and later time hadronic 
phase both have finite electrical conductivities, this finite conducting medium will definitely 
modify the decay of magnetic field according to the laws of magneto hydrodynamics
(MHD) ~\cite{Roy:2015kma,Pu:2016ayh,Hongo:2013cqa,Inghirami:2016iru,Inghirami:2019mkc}
or through a transport simulation~\cite{Das:2016cwd} , a matter 
which is still under investigation Refs.~\cite{Deng:2012pc,Satow:2014lia,Skokov:Illarionov}.  
Usually the transport coefficients such as shear, bulk viscosity, and electrical conductivity 
are taken as an input to dynamical models such as relativistic MHD. Hence it is important
to calculate these transport coefficients possibly the temperature dependence in presence of 
strong electro-magnetic fields from the underlying microscopic theories. The calculation of 
transport coefficients in quark and hadronic matter in presence of a magnetic field were carried out in recent 
Refs.~\cite{Tuchin,Li_shear,Asutosh,G_NJL_B,JD1,JD2,HM_TrB,Sedrakian_el,Kerbikov:2014ofa,Nam:2012sg,Huang:2011dc,Hattori:2016lqx,
Manu1,Manu2,Feng_cond,Fukushima_cond,Arpan1,Arpan2,NJLB_el,Hattori_bulk,Sedarkian_bulk,Agasian_bulk1,Agasian_bulk2,Manu3,
Denicol:2018rbw,Manu4,Balbeer,Balbeer1}, where shear viscosity~\cite{Tuchin,Li_shear,Asutosh,G_NJL_B,JD1,JD2,HM_TrB}, 
electrical conductivity~\cite{JD1,JD2,HM_TrB,Sedrakian_el,Kerbikov:2014ofa,
Nam:2012sg,Huang:2011dc,Hattori:2016lqx,Manu1,Manu2,Feng_cond,Fukushima_cond,Arpan1,Arpan2,NJLB_el}, and 
bulk viscosity~\cite{Hattori_bulk,Sedarkian_bulk,Agasian_bulk1,Agasian_bulk2,Manu3} were calculated 
in presence of a magnetic field.  
The dynamics of  heavy quark in presence of magnetic field within the framework of Fokker-Planck 
equation was studied in ~\cite{Manu4,Balbeer}.
In the present work, we carry out a similar investigation, where
we consider a multi-component Hadron Resonance Gas and 
evaluate the shear viscosity and electrical conductivity in the presence of a magnetic field.
In principle, one can calculate these transport coefficients in the presence 
of a magnetic field by solving QCD on a space-time lattice, but due to the current computational limitation 
and some technical difficulties it is unlikely to obtain the accurate result of 
these quantities in the low-temperature regime.
However, it is well known that HRG model successfully reproduces Lattice data just below the 
crossover temperature ($T_{c}$)~\cite{HRG_rev}
and it is expected that at much lower temperatures HRG as an effective model can be reliably used 
to calculate transport coefficients of hadronic matter. Since the magnetic field is 
non-zero in the hadronic phase it motivates us to calculate the transport coefficients in presence of the magnetic field. In refs {\cite{Endrodi,Subhasish}}, thermodynaimcal properties of hadron resonance gas in presence of magnetic field has been investigated. 

Here we would like to mention that recently in Refs.~\cite{HM_TrB,Arpan1,Arpan2} transport coefficients (electrical conductivity and shear viscosity) for a HRG were 
studied in presence of the magnetic field using the relaxation time approximation. The relaxation time was obtained 
from the constant cross section of hadrons. One of the crucial difference 
between the present work and the previous work Ref.~\cite{HM_TrB} is that we give a general framework of using projection 
tensors \cite{Hess} consisting of magnetic and hydrodynamical tensor degrees of freedom along which the 
viscous correction to single particle distribution function can be systematically expanded in a 
Chapman-Enskog (CE) series. This is unlike the heuristic basis \cite{Landau} used in the previous works. 
Hence, the present formalsim can be used to systematically derive second and higher order non-resistive 
MHD equations in the lines of Ref.~\cite{Denicol:2018rbw} but using a general CE series expansion. 
Apart from this important technical difference, in the present study we have calculated all the transport coefficients which are available in a Landau frame i.e., shear viscosity, bulk viscosity, and baryon diffusion (as well as electrical conductivity)
for hadronic matter. Additionaly, we do  not estimate the relaxation 
time from the hadronic size but rather treat this as a free parameter. 
In the present work we have separately explored the contributions of neutral and electrical charged hadrons to shear viscosity which might be important phenomenologically.
 Due to the Lorentz force the transport coefficients for electrically charged hadrons becomes anisotropic, whereas,
the neutral hadrons only contribute to the isotropic transport processes. We give some estimate of the relative contribution of such 
anisotropic and the isotropic transport coefficients within the HRG model for zero and non-zero $\mu_B$.
%In this regards, present work has attempted similar topics but gone through a different sketch of estimations, which are
%missing in earlier references. First of all, instead of going hard sphere calculation of relaxation time, which is
%never be exact mapping of hadron interaction, we have taken it as a parameter. With proper normalization of it,
%we have presented the phase-space of transport coefficients. In absence of magnetic field, this normalization
%draw similar kind pattern as noticed for thermodynamical quantities like pressure, energy density, entropy density,
%when one goes from non-interacting to interacting picture via massless QGP to HRG results. In presence of magnetic
%field, an anisotropic phase-space of transport coefficients can be obtained. Present article have attempted
%to sketch that temperature and magnetic field dependent anisotropic phase-space, carried by different transport
%coefficients.

The article is organized as follows: in Sec.~\ref{sec:Formalism}, we briefly discuss the thermodynamics  of the HRG model.  
In Sec.~\ref{sec:BTE} we introduce the Boltzmann transport equation in relaxation time approximation and the ansatz for 
the off-equilibrium distribution function required to calculate the transport coefficients.  
In the same section we discuss the transport coefficients obtained from relaxation time approximation with and without the 
magnetic field. Next, in Sec.~\ref{sec:res} we discuss numerical results obtained for HRG. We give a summary of our work in Sec.~(\ref{sec:sum})
At the end detailed derivation of various transport coefficients are given in Appendix. Throughout the paper we use the natural unit, the four vectors are denoted by the greek indices and three vectors are denoted by the latin indices unless stated otherwise.

% Next in Sec.~(\ref{sec:th_TB}), the mapping of LQCD
% thermodynamics in presence of magnetic field has been addressed, where two different
% methodologies are classified into two subsections - (\ref{sec:gTB}) and (\ref{sec:ZTB}).
% After developing the quasi particle description, it is applied to estimate transport
% coefficients like shear viscosity and electrical conductivity at finite magnetic 
% field in Sec.~(\ref{sec:sh_el}), whose framework is addressed in Sec.~(\ref{sec:App}) with
% two subsections - (\ref{sec:Sh_B}) and (\ref{sec:el_B}). At the end in Sec.~(\ref{sec:sum})
% we have summarized our investigations.

\section{Formalism}
\label{sec:Formalism}
\subsection{Thermodynamics}
Here, we start with a brief discussion of the hadron resonance gas (HRG) model %description for different
to define the thermodynamical quantities like entropy density $s$, enthalpy per particle $h$, etc. which are
used for the calculations of different transport coefficients. %Our starting point will
All thermodynamic quantities are derived from the grand canonical partition function $Z$ of the hadronic 
matter with volume $V$ at temperature $T$ and chemical potential of $i^{th}$ species $\mu_i$:
\be
{\rm ln}Z=V\sum_{i}\int \frac{d^3{\vec p_i}}{(2\pi)^3}g_i r_i{\rm ln}\Big[1+r_{i} e^{\beta(p^{0}_i-\mu_i)}\Big]~,
\ee
where, $\mu_i=B_i \mu_B$ with $B_i$ as the baryon number of the hadronic species,
$\mu_B$ as baryon chemical potential. $g_i$, $p^{0}_i=\{{\vec{p_i}^2}+m_i^2\}^{1/2}$ are degeneracy factors and energy of the hadrons of species 
$i$ with mass $m_i$; $r_i=\pm$ stands for 
fermion or bosons respectively. The total degeneracy factor of a particular species of hadron is obtained as
$g_i=g^s_i\times g^I_i$, where $g^s_i$, $g^I_i$ are the spin and iso-spin degeneracy factors respectively.

Once the partition function is defined, the thermodynamic quantities pressure ($P$), energy density 
($\epsilon$), net baryon density ($\rho$) are calculated  from the following standard definitions:
\bea
P&=& \frac{T}{V}{\rm ln}Z,
\nn\\
\epsilon &=& \frac{T^2}{V}\frac{\del}{\del T}{\rm ln}Z,
\nn\\
\rho &=& \frac{T}{V}\frac{\del}{\del \mu}{\rm ln}Z~.
\label{Pen}
\eea
Using Eqs.~(\ref{Pen}), we can further define the entropy density $s$, and the enthalpy per particle $h$
by using the relations
\bea
s &=& \sum_{i} (\epsilon +P -\mu_i \rho_i)/T,
\nn\\
h &=& (\epsilon +P)/\rho.\label{Eq:h}
\eea
Where, $\rho_i$ is baryon density of hadron species $i$.

%%%%%%%%%%%%%%%%% Boltzmann transport equation %%%%%%%%%%%%%%%%%%%%%
\section{Boltzmann transport equation}
\label{sec:BTE}
The calculation of all the transport coefficients considered here are based on relaxation time approximation of the collision
kernel of the Boltzmann equation, hence, it is worthwhile to discuss the method for the sake of completeness.
The general form of the Boltzmann equation in the presence of external fields, in the relaxation time approximation
is given by ~\cite{Landau,Asutosh,JD1,JD2},
\be
p^{\mu}\partial_{\mu}f_i + qF^{\mu \nu}p_\nu \frac{\partial f_i}{\partial p^\mu} = -\frac{U\cdot p}{\tau_{c}}\delta f_i~,
\label{RBE1}
\ee
where, $F^{\mu \nu}$ is the electromagnetic field strength tensor. 
%and for comparetively very weak electric field 
For our case, only magnetic field is present, hence
$F^{\mu \nu} =-B^{\mu \nu}$ with $B^{\mu\nu}=\epsilon^{\mu\nu\rho\alpha}B_\rho U_{\alpha}$. $B$ is the magnetic 
field strength and $b^\mu$ is the unit four vector defined as $b^{\mu}=\frac{B^{\mu}}{B}$. So, for a small deviation of the distribution function 
from the equilibrium, Eq.~(\ref{RBE1}) can be written as,
\be
p^{\mu}\partial_{\mu}f_{i0} = \Big(-\frac{U\cdot p}{\tau_{c}}\Big)\Big[1-\frac{qB \tau_{c}}{U\cdot p}b^{\mu\nu}p_{\nu}
\frac{\partial}{\partial p^{\mu}}\Big]\delta f_i~.
\label{RBET}
\ee
The equilibrium distribution function for $i^{th}$ hadron species is $f_{i0}=(e^{\beta(U\cdot p - \mu_i)}+r)^{-1}$, where $r=\pm 1$ depending on the statistics. In all proceeding calculations, hydrodynamic four-velocity $u^\mu$ is defined in the Landau frame such that $u_\nu T^{\mu\nu} = \sum_i \int d^3p p^{\mu}_i f_i = \epsilon u^\mu$, where $T^{\mu\nu}$ is energy-momentum tensor and $\epsilon$ is the energy density.

Here we construct $\delta f_i$  as a linear combination of the thermodynamic forces times appropriate tensorial coefficients so that
$\delta f_i$ turns out to be a Lorentz scalar, 
\be
\delta f_i=A_iX+ B_i^{\mu}X_{\mu} + C_i^{\mu\nu}X_{\mu\nu}.
\ee
Where $X_{\mu\nu...}$ represents the thermodynamic forces. Replacing the above form of $\delta f_i$ in the Boltzmann transport equation and comparing the coefficients of the 
thermodynamic forces we get the unknown coefficients $A_i,B_i^{\mu}$ and $C_i^{\mu\nu}$ in the expression for $\delta f_i$. Using the $\delta f_i$ in the thermodynamic 
flows we obtain the transport coefficients as discussed in details in the Appendix \ref{sec:App}.

Subsequently the the dissipative quantities like current density $(J^{\mu}_D)$, stress tensor $(\pi^{\mu\nu})$, bulk viscous pressure $(\Pi)$, and particle diffusion current $(n^\mu)$ can be written as:\\

\begin{equation}
\begin{aligned}
J^\mu_D &=\sigma^{\mu\nu}E_{\nu}\\
\pi^{\mu \nu} &=\eta^{\mu\nu\alpha\beta}V_{\alpha\beta}\\
\Pi &=\zeta^{\mu\nu}\partial_\mu u_\nu\\
n^\mu &=\kappa^{\mu\nu}\nabla_\nu(\mu/T)
\end{aligned}
\end{equation}
where the tensor coefficients $\sigma_{\mu\nu}$ is given in Eqn.~(\ref{Eq:deltafCond2}) and the rest can be written as
\begin{equation}
\begin{aligned}
\eta^{\mu\nu\alpha\beta}&=\frac{1}{15}\sum_i g_i\int  \frac{d^3 p_i{(\vec{p_i})}^4}{{(2\pi)^3}p_i^0}C_i^{(n)\mu\nu\alpha\beta}\\
\zeta^{\mu\nu}&=\frac{1}{3}\sum_i g_i\int \frac{d^3 p_i {(\vec{p_i})}^2}{{(2\pi)^3}p_i^0} C_i^{(n)\mu\nu}, \\
\kappa^{\mu\nu}&=-\frac{1}{3}\sum_{i}^\text{baryons} g_i\int \frac{d^3 p_i {(\vec{p_i})}^2}{{(2\pi)^3}p_i^0} K_i^{(n)\mu\nu}
\end{aligned}
\end{equation}
where the coefficients $C^{(n)\mu\nu\alpha\beta},C^{(n)\mu\nu}$ and $K^{(n)\mu\nu}$ are given in in Eqns.~(\ref{Eq:deltafShear}, \ref{Eq:deltafBulk}, \ref{Eq:deltafDiff})  respectively. Note that here diffusion current refers to baryon diffusion and hence the sum is over all baryons (anti-baryons).
\subsection{Transport coefficients without a magnetic field}
After the short discussion on the thermodynamical quantities, we discuss here about
the transport coefficients of a relativistic systems of particles in absence of any external magnetic fields.
The electrical conductivity ($\sigma$), 
shear viscosity ($\eta$), bulk viscosity ($\zeta$) and the diffusion coefficient ($\kappa$) for a HRG are given in terms of the temperature and the relaxation time of hadrons,
\bea
\sigma &=& \sum_i g^s_iq_i^2 \frac{1}{3T}\int \frac{d^3{ p_{i}}}{(2\pi)^3}\frac{|\vp_{i}|^2}{(p_{i}^{0})^2}\tau_c f_{i0}(1-r_i f_{i0})
\nn\\
\eta &=& \sum_i \frac{g_i}{15T}\int\frac{d^{3}{\vec p_{i}}}{(2\pi)^{3}}\frac{|\vp_{i}|^4}{(p_{i}^{0})^2}\tau_{c}f_{i0}(1-r_if_{i0})
\nn\\
\zeta &=&\sum_{i}\frac{g_{i}}{T}\int \frac{d^{3}{\vec p_{i}}}{(2\pi)^{3}(p_{i}^{0})^{2}}Q_{i}^{2}\tau_{c} f_{i0}(1-r_{i}f_{i0}).
\nn\\
\kappa &=& \sum_i \frac{g_i}{3h}\int  \frac{d^{3}{\vec p_{i}}}{(2\pi)^{3}}
\frac{|\vp_{i}|^2}{(p_{i}^{0})^2}\tau_c(h-p_{i}^{0})f_{i0}(1-r_if_{i0})~,
\nn\\
%\sigma &=& \sum_h g^s_hq_h^2 \frac{\beta}{3}\int \frac{d^3\vp}{(2\pi)^3}\frac{\vp^2}{\om_h^2}\tau_c f_0(1-r_h f_0)
%\nn\\
%\eta &=& \sum_h \frac{g_h}{15T}\int\frac{d^{3}\vp}{(2\pi)^{3}}\frac{\vp^4}{\om_h^2}\tau_{c}f_{0}(1-r_hf_{0})
%\nn\\
%\kappa &=& \sum_h \frac{g_h}{3h}\int  \frac{d^{3}\vp}{(2\pi)^{3}}\frac{\vp^2}{\om_h^2}\tau_c(h-\om_h)f_{0}(1-r_hf_{0})~,
%\nn\\
\label{Tr_B0}
\eea
where $q_i$ stands for electric charge of hadrons type $i$ , $\tau_c$ is the relaxation time of hadrons, 
which is taken to be same for all hadrons for the sake of simplicity. The $Q_i$ is a function of speed of 
sound along with other thermodynamic quantities the details of which  is given in Appendix \ref{sec:App}. 
The derivation of the transport coefficients given in Eq.(\ref{Tr_B0}) can be found in Refs.~\cite{Gavin,Purnendu} 
as well as in  the Appendix \ref{sec:App}. Similar expressions can also be obtained in Kubo relation~\cite{G_IJMPA,SG_PRD}. 

%Present article
%is aimed for transport coefficients of HRG system in presence of magnetic field. Hence Appendix is mainly
%dedicated for finite magnetic field expressions of different transport coefficients, where without magnetic
%field expressions of transport coefficients will already be there. 
In the present article, we aim to calculate the transport coefficients of HRG in presence of a magnetic field, 
the values of these coefficients 
without the magnetic fields are obtained by taking the limit of vanishing magnetic field. 
The expression for the transport coefficients in the presence of magnetic fields are given in 
the next few sub-sections and the corresponding detailed derivation  for the 
same is given in Appendix \ref{sec:App}.

\subsection{Electrical conductivity in magnetic field}
In presence of a magnetic field, the transport coefficients involve another time scale, cyclotron  time
$\tau_{iB}= p_i^{0}/(eB)$  along with the usual  relaxation time $\tau_c$ which usually depends on the rate of contact collisions between the constituents. The index $i$ refers to type of hadronic species.

The non-zero Lorentz force due to the magnetic fields give rise to an anisotropic
transport phenomenon (as the force along the magnetic field is zero and non zero in other directions), it is obvious that if the collision time $\tau_c$ is much smaller  the cyclotron time $\tau_{iB}$ the effect of magnetic field is negligible i.e., the system is almost isotropic when $\tau_c/\tau_{iB}  \ll 1$, and  it becomes anisotropic when  $\tau_c/\tau_{iB}  \sim 1$ or greater . We also note that along the magnetic field, Lorentz force does not work, so the parallel component of any transport coefficient (denoted by $\parallel$) remains the same as without the magnetic field, given in Eqs.~(\ref{Tr_B0}).  Here we need a little bit more clarification, in linear theory, any thermodynamic fluxes are proportional to the corresponding thermodynamic forces and the proportionality constants are known as transport coefficients. If the system is isotropic the transport coefficients are scalar, but for an anisotropic medium, the transport coefficients are components of a tensor. The decompositions of  the transport coefficient tensor in terms of the available basis ($u^{\mu}, g^{\mu\nu},b^{\mu}, b^{\mu\nu}$) are not unique and we choose here a particular combination such that the decomposition has a component parallel  to the magnetic field which is denoted with a subscript $\parallel$. 
Whereas, the remaining components can have two  or more components, usually denoted with a subscript $\perp$ and $\times$. 
The $\times$-component is basically Hall component, which was absent for $B=0$, while $\perp$-component
at $B=0$ will still exist and it will be exactly equal to $\parallel$-component, which restore the isotropic
property of the medium at $B=0$.
For electrical conductivity, the expressions of parallel ($\sigma_\parallel$),
perpendicular ($\sigma_\perp$) and cross ($\sigma_{\times}$) components for hadron resonance gas are given below

%Whereas, the plane, perpendicular to the magnetic field, can have two more components, marked as $\perp$ and $\times$.  For electrical conductivity, the expressions of parallel ($\sigma_\parallel$), perpendicular ($\sigma_\perp$) and cross ($\sigma_{\times}$) components for HRG system are given below
\bea
\sigma_\parallel&=&\sum_{i} g^s_iq_i^2 \frac{\beta}{3}\int 
\frac{d^3{\vec p_{i}}}{(2\pi)^3}\frac{|\vp_i|^{2}}{(p^{0}_{i})^2}\tau_c f_{i0}(1-r_i f_{i0}),
\nn\\
\sigma_\perp&=&\sum_i g_iq_i^2 \frac{\beta}{3}\int 
\frac{d^3{\vec p_{i}}}{(2\pi)^3}\frac{|\vp_i|^{2}}{(p^{0}_{i})^2}
\frac{\tau_c}{1+(\tau_c/\tau_{iB})^2} f_{i0}(1-r_i f_{i0}),
\nn\\
\sigma_{\times}&=&\sum_i g_i q_i^2 \frac{\beta}{3}\int 
\frac{d^3{\vec p_{i}}}{(2\pi)^3}\frac{|\vp_i|^{2}}{(p^{0}_{i})^2}
\frac{(\tau_c^2/\tau_{iB})}{1+(\tau_c/\tau_{iB})^2} f_{i0}(1-r_i f_{i0}).
\nn\\
%\sigma_\parallel&=&\sum_{h} g^s_hq_h^2 \frac{\beta}{3}\int \frac{d^3\vp}{(2\pi)^3}\frac{\vp^2}{\om_h^2}\tau_c f_0(1-r_h f_0),
%\nn\\
%\sigma_\perp&=&\sum_h g_hq_h^2 \frac{\beta}{3}\int \frac{d^3\vp}{(2\pi)^3}\frac{\vp^2}{\om_h^2}\tau_c
%\frac{1}{1+(\tau_c/\tau^h_B)^2} f_0(1-rf_0),
%\nn\\
%\sigma_{\times}&=&\sum_h g_h q_h^2 \frac{\beta}{3}\int \frac{d^3\vp}{(2\pi)^3}\frac{\vp^2}{\om_h^2}\tau_c~
%\frac{(\tau_c/\tau^h_B)}{1+(\tau_c/\tau^h_B)^2} f_0(1-r_h f_0).
%\nn\\
\label{sig_b}
\eea
As mentioned earlier, the detail derivation is given in Appendix  \ref{sec:AppB}. To compare our results for electrical 
conductivities Eq.(\ref{sig_b}) to some of the earlier findings Refs.~\cite{Sedrakian_el,JD1,JD2}, where the conductivities 
are denoted with $\sigma_{0,1,2}$ , we found the following relations holds 
\bea
\sigma_\parallel &=&\sigma_0 +\sigma_2,
\nn\\
\sigma_\perp &=& \sigma_0,
\nn\\
\sigma_\times &=& \sigma_1~.
\eea

%%%%%%%%%%%%%%%%% Shear Viscosity %%%%%%%%%%%%%%%%%
\subsection{Shear Viscosity in magnetic field}
The most general form of the $\delta f_i $ in presence of a magnetic field where only shear stress is present is given by,
\bea
\delta f_i&=&\sum\limits_{n=0}^{4} c_{n}C^{(n)}_{\mu\nu\alpha\beta}p_i^{\mu}p_i^{\nu}V^{\alpha\beta}\\
&=&\Big[ c_{0} P^{0}_{\langle \mu\nu\rangle\alpha\beta}+ c_{1}\big(P^{1}_{\langle \mu\nu\rangle\alpha\beta}+P^{-1}_{\langle \mu\nu\rangle\alpha\beta}\big) \nonumber\\
&+& ic_{2}\big(P^{1}_{\langle \mu\nu\rangle\alpha\beta}-P^{-1}_{\langle \mu\nu\rangle\alpha\beta}\big) + c_{3}\big(P^{2}_{\langle \mu\nu\rangle\alpha\beta}+P^{-2}_{\langle \mu\nu\rangle\alpha\beta}\big)\nonumber\\
&+& ic_{4}\big(P^{2}_{\langle \mu\nu\rangle\alpha\beta}-P^{-2}_{\langle \mu\nu\rangle\alpha\beta}\big) \Big]p_i^{\mu}p_i^{\nu}V^{\alpha\beta}~,
\eea
where $V_{\alpha \beta} = \frac{1}{2}(\frac{\partial U_{\alpha}}{\partial x_{\beta}} 
+ \frac{\partial U_{\beta}}{\partial x_{\alpha}})$, the form of projectors $P^{n}_{\langle \mu\nu\rangle\alpha\beta}$ will be given in Appendix \ref{sec:App} for $n=-2,-1,0,1,2$. Using this 
expression for $\delta f_i$, the shear viscous coefficients turnout to be,
\iffalse
\bea
\eta_{\parallel}=\sum_{i}\frac{2g_{i}}{15}\int\frac{d^{3}{\vec p_{i}}}{(2\pi)^{3}p_{i0}}|\vp_i|^{4}c_{i0}, \label{eq:shear1} \\
\eta_{\perp}=\sum_{i}\frac{2g_{i}}{15}\int\frac{d^{3}{\vec p_{i}}}{(2\pi)^{3}p_{i0}}|\vp_i|^{4}c_{i1},\\
\eta'_{\perp}=\sum_{i}\frac{2g_{i}}{15}\int\frac{d^{3}{\vec p_{i}}}{(2\pi)^{3}p_{i0}}|\vp_i|^{4}c_{i3},\\
\eta_{\times}=\sum_{i}\frac{2g_{i}}{15}\int\frac{d^{3}{\vec p_{i}}}{(2\pi)^{3}p_{i0}}|\vp_i|^{4}c_{i2},\\
\eta'_{\times}=\sum_{i}\frac{2g_{i}}{15}\int\frac{d^{3}{\vec p_{i}}}{(2\pi)^{3}p_{i0}}|\vp_i|^{4}c_{i4}.
\label{eq:shear5}
\eea
\fi
\bea
\eta_{\parallel}&=&\sum_i\frac{g_i}{15T}\int\frac{d^3{\vec p_i}}{(2\pi)^{3}}\frac{|\vp_i|^4}{p_{i0}^2} \tau_{c} f_{i0}(1-r_if_{i0}) 
\nn\\
\eta_{\perp}&=&\sum_i\frac{g_i}{15T}\int\frac{d^3{\vec p_i}}{(2\pi)^{3}}\frac{|\vp_i|^4}{p_{i0}^2}\frac{\tau_{c}}{1+(\tau_c/\tau_{iB})^2} f_{i0}(1-r_if_{i0}) 
\nn\\
\eta'_{\perp}&=&\sum_i\frac{g_i}{15T}\int\frac{d^3{\vec p_i}}{(2\pi)^{3}}\frac{|\vp_i|^4}{p_{i0}^2} \frac{\tau_{c}}{1+(2\tau_c/\tau_{iB})^2} f_{i0}(1-r_if_{i0}) 
\nn\\
\eta_{\times}&=&\sum_i\frac{g_i}{15T}\int\frac{d^3{\vec p_i}}{(2\pi)^{3}}\frac{|\vp_i|^4}{p_{i0}^2}
\frac{\tau_c^2/\tau_{iB}}{1+(\tau_c/\tau_{iB})^2} f_{i0}(1-r_if_{i0}) 
\nn\\
\eta'_{\times}&=&\sum_i\frac{g_i}{15T}\int\frac{d^3{\vec p_i}}{(2\pi)^{3}}\frac{|\vp_i|^4}{p_{i0}^2}\frac{\tau_c^2/\tau_{iB}}{\frac{1}{2}
+2(\tau_c/\tau_{iB})^2}f_{i0}(1-r_if_{i0})\nn\\\label{eq:shear5}
\eea
The coefficients $\eta_{\parallel}$, $\eta_{\perp}$, $\eta'_{\perp}$ are even functions of magnetic field $B$. The two coefficients $\eta_{\times}$,
 $\eta'_{\times}$ may have either sign and they are odd functions of $B$. The later two coefficients are also called transverse viscosity coefficients \cite{Hess}. 
 %fully transverse, fully hall, and ordinary transverse and ordinary Hall viscosity}. 
 %components  are in the  $\perp$-direction {\color{red} to the $B^{\mu}$},
%two Hall components $$, $\eta'_{\times}$ in $\times$-direction and one $\eta_{\parallel}$ in
%$\parallel$-direction. {\color {red} 
We note the expressions for shear viscosities given in Eqns.(\ref{eq:shear5})  are identical to those given in Refs.~\cite{Asutosh,JD1,JD2,Landau}.
%\bea
%\eta_0 &=& \eta_{\parallel}
%\nn\\
%\eta_2&=&\eta_{\perp}
%\nn\\
%\eta_4&=&\eta_{\times}
%\nn\\
%\eta_1&=&\eta'_{\perp}
%\nn\\
%\eta_3&=&\eta'_{\times}~.
%\eea

%%%%%%%%%%%%%%%%%%%%%%%%%%%%%%%%%%%%%%%%%%%%%%%%%%%

% %%%%%%%%%%%%%%%%% Bulk Viscosity %%%%%%%%%%%%%%%%%%
 \subsection{Bulk Viscosity in magnetic field}
Similarly for bulk viscosity we restrict ourselves to only the divergence of the fluid four velocity and neglect the other thermodynamic forces,
\bea
\delta f_i=\sum_{n=1}^{3}c_{n}C^{\mu\nu}_{n}\partial_{\mu}U_{\nu}.
\eea 
Using this $\delta fi$ the bulk viscous coefficients turns out to be,
\bea
\zeta_{\parallel}=\zeta_{\perp}&=&\sum_{i}\frac{g_{i}\tau_c}{T}\int \frac{d^{3}{\vec p_{i}}}{(2\pi)^{3}p_{i0}^{2}}
Q_{i}^{2}f_{i0}(1-r_{i}f_{i0}),\\
\zeta_{\times}&=&0.
\label{eq:zeta123}
\eea
The bulk viscous coefficients remains unchanged under the influence of the magnetic field as was 
also shown in Ref.~\cite{Denicol:2018rbw} using Grad's 14 moment approximation. 
The detailed derivation of Eq.(\ref{eq:zeta123}) is given in the Appendix \ref{sec:App}.
% 
% \bea
% \zeta_{\parallel}=\zeta_{\perp}&=&\frac{\tau_c}{T}\int \frac{d^3\bp}{(2\pi)^{3}p_{0}^{2}}Q^{2}f_{0}(1-rf_{0})
% \nn\\
% \zeta_{\times}&=&0
% \label{zeta_3}
% \eea

%%%%%%%%%%%%%%%%% Diffusion coefficients %%%%%%%%%%%%%%%%
\subsection{Net baryon diffusion coefficient in magnetic field}
For the case of diffusion we keep only the term containing the spacial derivative 
of $\mu /T$ in the expression for $\delta f_i$,
\bea
\delta f_i=K^{\mu\nu}p_{i\mu}\partial_{\nu}\Big(\mu_i/T\Big). 
\eea 
Using this $\delta f_i $ the diffusion coefficients turn out to be,
%Thermal diffusion coefficients in presence of magnetic field follows similar
%decomposition like the electric conductivity.
%Only electric charge flow will be replaced by flow difference between the energy and the enthalpy.
%Based on the derivation, given in Appendix \ref{sec:App}. Its three components for HRG system will be

\bea
\kappa_{\parallel}&=& \sum_{i}^\text{baryons}\frac{g_{i}}{3h}\int  \frac{d^3 {\vec p_{i}}}{(2\pi)^{3}}
\frac{|\vp_i|^2}{(p_{i}^{0})^2}\tau_c(h-B_ip_{i0})f_{i0}(1-r_{i}f_{i0})\nonumber
\\
\kappa_{\perp}&=& \sum_{i}^\text{baryons}\frac{g_{i}}{3h}\int  \frac{d^3 {\vec p_i}}{(2\pi)^{3}}
\frac{|\vp_i|^2}{(p^{0}_{i})^2}\frac{\tau_c(h-B_ip_{i0})}{1+(\frac{\tau_{c}}{\tau_{iB}})^2}f_{i0}(1-r_{i}f_{i0})\nonumber
\\
\kappa_{\times}&=& \sum_{i}^\text{baryons}\frac{g_{i}}{3h}\int  \frac{d^3 {\vec p_i}}{(2\pi)^{3}}
\frac{|\vp_i|^2}{(p^{0}_{i})^2}\frac{\tau_c(\frac{\tau_{c}}{\tau_{iB}})(h-B_ip_{i0})}
{1+(\frac{\tau_{c}}{\tau_{iB}})^2}f_{i0}(1-r_{i}f_{i0}),\nonumber\\
\label{Diff_3}
\eea
where $h$ is the enthalpy density as defined in Eq.~(\ref{Eq:h}) and the sum runs over baryons only. Due to the anisotropy induced by
the magnetic field we have three diffusion coefficients. Here again, the details can be found in Appendix~\ref{sec:App}.

\section{Results}
\label{sec:res}
\begin{figure} %{ht}
\centering
\includegraphics[scale=0.45]{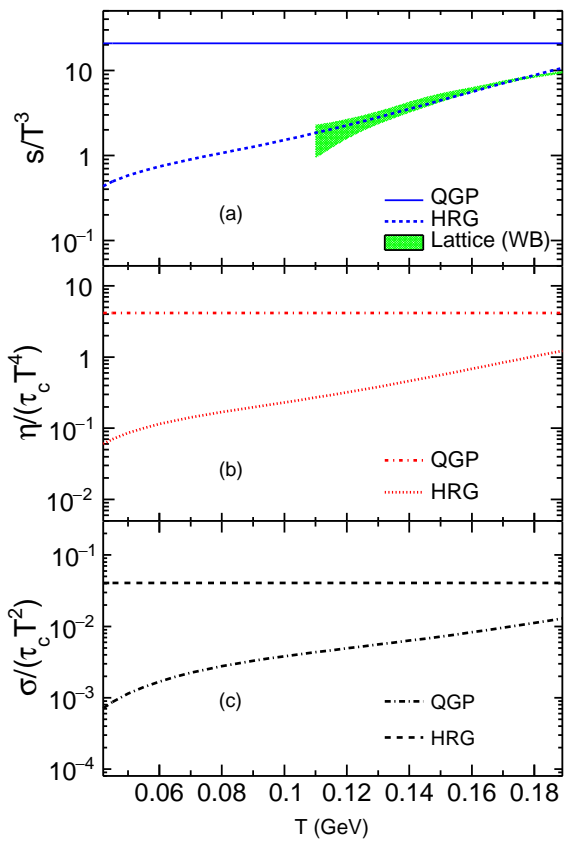}
\caption{(a) Normalized entropy density $s/T^3$, (b) shear viscosity $\eta/(\tau_cT^4)$
and (c) electrical conductivity $\sigma/(\tau_c T^2)$ as function of $T$ for massless QGP (horizontal lines)
and HRG.
}
\label{fig:s_Tr_T}
\end{figure}

In the formalism section, we have summarized the analytic expressions for the anisotropic components of
the shear viscosity, bulk viscosity, thermal diffusion and the electrical conductivity for a finite magnetic field.
In this section, we will explore the temperature and magnetic field dependence of these transport coefficients
 for HRG model calculations. 

\begin{figure*} %{ht}
\centering
\includegraphics[width=0.48\textwidth]{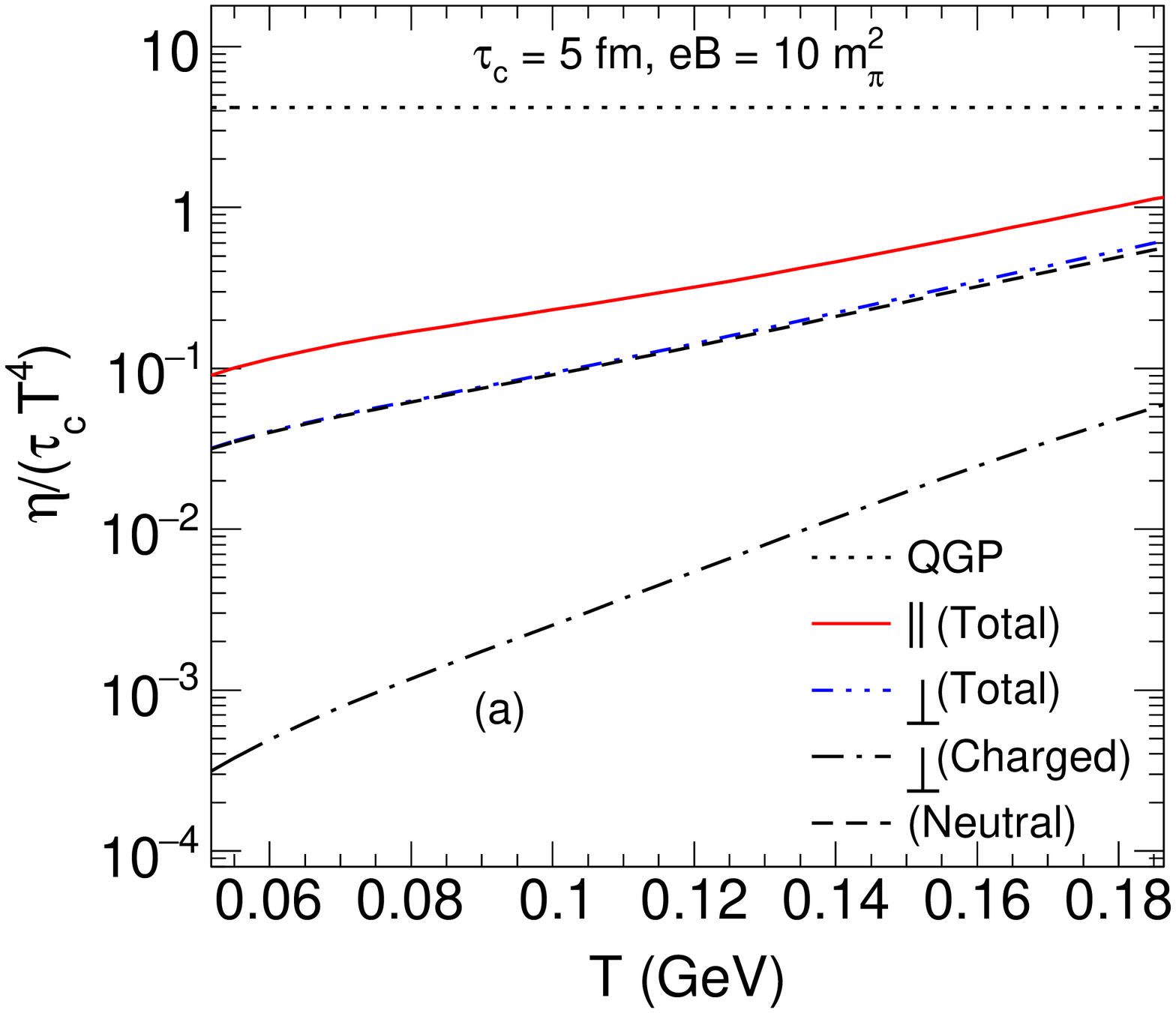}
\includegraphics[width=0.48\textwidth]{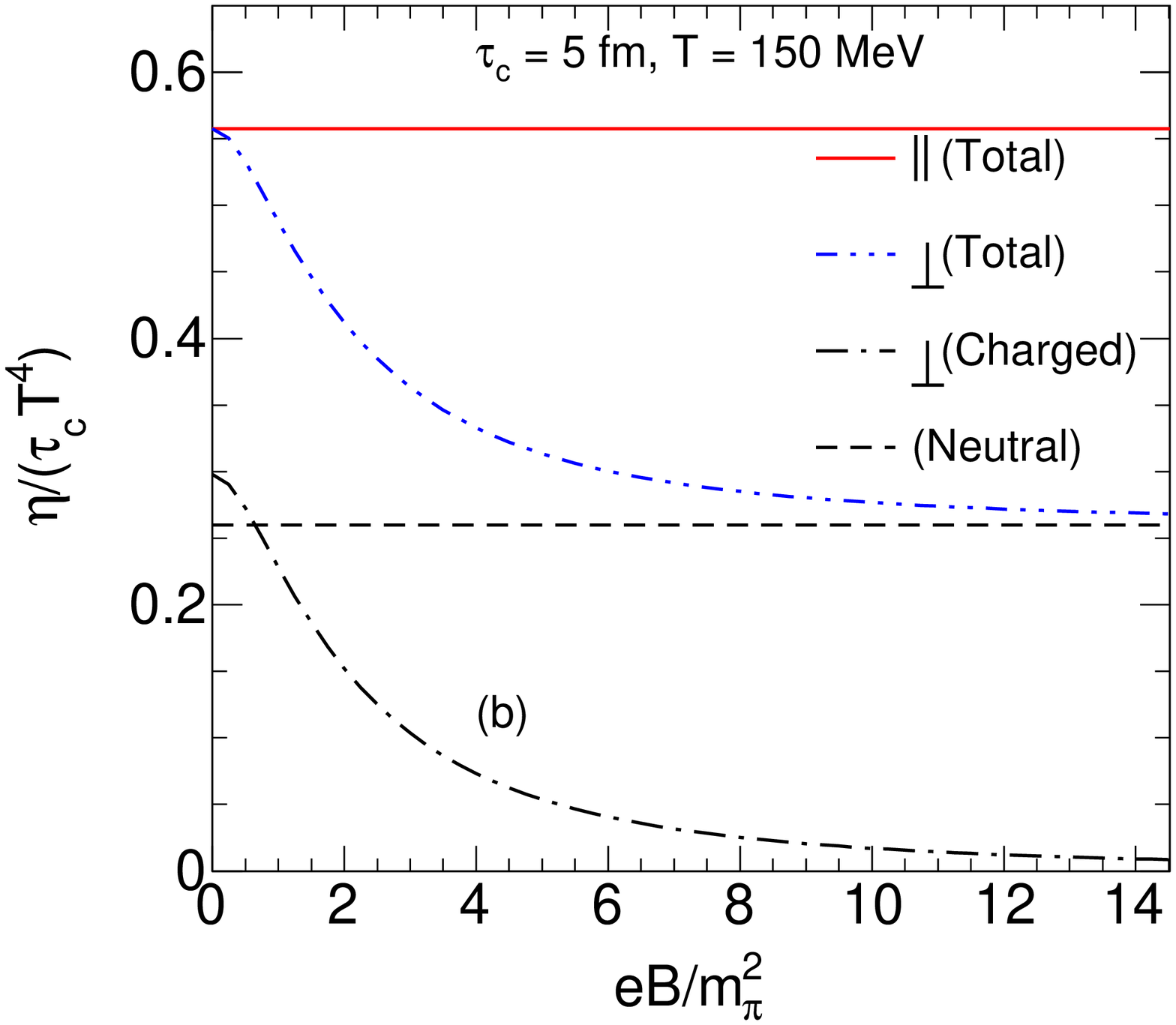}
\caption{The anisotropic component of the shear viscosities $\eta_{\bot}$, $\eta_{\parallel}$ for HRG and isotropic value for massless QGP are plotted against the axes of
(a)temperature ($T$) of the medium, (b) external magnetic field ($B$).
%(c) tharmal relaxation time $\tau_c$ in the framework of HRG model.
}
\label{fig:eta_TB}
\end{figure*}

Before discussing the results for HRG with physical masses of hadrons let us first consider the
simpler massless case for quark gluon plasma (QGP).
%simpler case of a conformal HRG where all the rest  masses are set to zero. 
Here we also 
compare the result obtained from our numerical implementation of the HRG model to that of a Lattice QCD (LQCD)
result for a sanity check.
%For getting reference points of our numerical results, let us first explore
%the massless results of transport coefficients. 
In the massless limit (also known as the Stefan-Boltzmann (SB) limits) the thermodynamical quantities like Pressure ($P$), energy density ($\epsilon$), 
varies as $T^{4}$ and the entropy density ($s $) varies as $T^{3}$, more explicitly  
\bea
P_{SB} &=&g\frac{\zeta(4)}{\pi^2}T^4, % = g\frac{\pi^2}{90}T^4
\nn\\
\epsilon_{SB} &=& g\frac{3 \zeta(4)}{\pi^2}T^4, %= g\frac{3 \pi^2}{90}T^4
\nn\\
s_{SB} &=& g\frac{4 \zeta(4)}{\pi^2}T^3~,
%s &=& (\epsilon + P)/T
%\nn\\
%&=& g\frac{4 \zeta(4)}{\pi^2}T^3 = g\frac{4 \pi^2}{90}T^3~,
\eea
where $\zeta(4)$ stands for zeta function.
Here the subscript $SB$ stands for the Stefan-Boltzmann (SB) limit, in this limit the interaction measure $(\epsilon-3P)/T^{4}$ becomes zero and we consider the HRG to be a non-interacting gas. %or non-interacting limits.
It is clear that in the SB limit $P/T^{4}, \epsilon/T^{4}$, and $s/T^{3}$ are constants for a given degeneracy. For example, a 3 flavor quark-gluon-plasma with the degeneracy factor $g=16+\frac{7}{8}(24+12)=47.5$  yields $P/T^4=5.2$, $\epsilon/T^4=15.6$ and $s/T^3=20.8$.
However, for the physical masses of hadrons all these thermodynamics quantities have a smaller value than their 
corresponding SB values and approaches SB value from below as $m/T \rightarrow \infty$.
This is shown in the top panel of Fig.~\ref{fig:s_Tr_T}(a) for the normalized entropy density, where the result obtained from HRG model is shown by the blue dotted line, the corresponding $s_{SB}/T^3$ is shown by the blue horizontal solid line. 
For comparison we also show the LQCD (shown by green band) result  from Ref.~\cite{Borsanyi:2013bia} in the temperature range 120-180 MeV. 
It is clear from the Fig.~\ref{fig:s_Tr_T}(a)  that the normalized entropy density obtained from the Lattice QCD calculation and HRG matches very well in the temperature range considered here, also both results approaches 
SB value as temperature increases. 

Now let us discuss the shear viscosity and the electrical conductivity of a massless gas without 
any magnetic field as given in Eqs.~(\ref{Tr_B0}). In the massless limit the corresponding 
expressions are~\cite{JD2} :
\bea
\eta &=& g\frac{4\zeta(4)}{5\pi^2}\tau_c T^4,
\nn\\
\sigma &=& g_q q^2\frac{\zeta(2)}{3\pi^2}\tau_c T^2~,
\eea
where $g_q q^2=12\times\Big(\frac{4e^2}{9}+\frac{e^2}{9}+\frac{e^2}{9}\Big)=8e^2$ for 3 flavor QGP.
 We note that similar to the thermodynamic quantities, in the SB limit, the normalized shear viscosity and electrical conductivity $\eta/(\tau_c T^4)$ and $\sigma/(\tau_c T^2)$ are constants only depends on the degeneracy factor. 
These normalized SB values $\eta_{SB}/(\tau_c T^4$) and $\sigma_{SB}/(\tau_c T^2)$ are shown by the red dash-dotted and black dash horizontal lines in Fig.~\ref{fig:s_Tr_T}(b) and (c) respectively.
For an HRG both $\eta/(\tau_c T^4)$ and $\sigma/(\tau_c T^2)$ %when Eq.~(\ref{Tr_B0}) we get 
has smaller values compared to their corresponding SB values and approaches to SB value from below in the large temperature limit as shown by the red dot and black dash-dotted lines in Fig.~\ref{fig:s_Tr_T}(b) and (c) respectively. 

The striking similarity between the temperature dependence of thermodynamic quantity $s/T^{3}$ and
the transport coefficients, like $\eta/(\tau_c T^4)$, $\sigma/(\tau_c T^2)$ clearly shows that we may gain information
about the degrees of freedom of the system under consideration. Alternatively, we might get information about relaxation time $\tau_c$ if the temperature dependence of $\eta$ and $\sigma$ are known from other means. 

%Hence, at a glance suppressed values of thermodynamical quantities, like $s/T^3$
%are trying to mimic QCD interaction through thermodynamical phase-space.%
\begin{figure*} %{ht}
\centering
\includegraphics[width=0.48\textwidth]{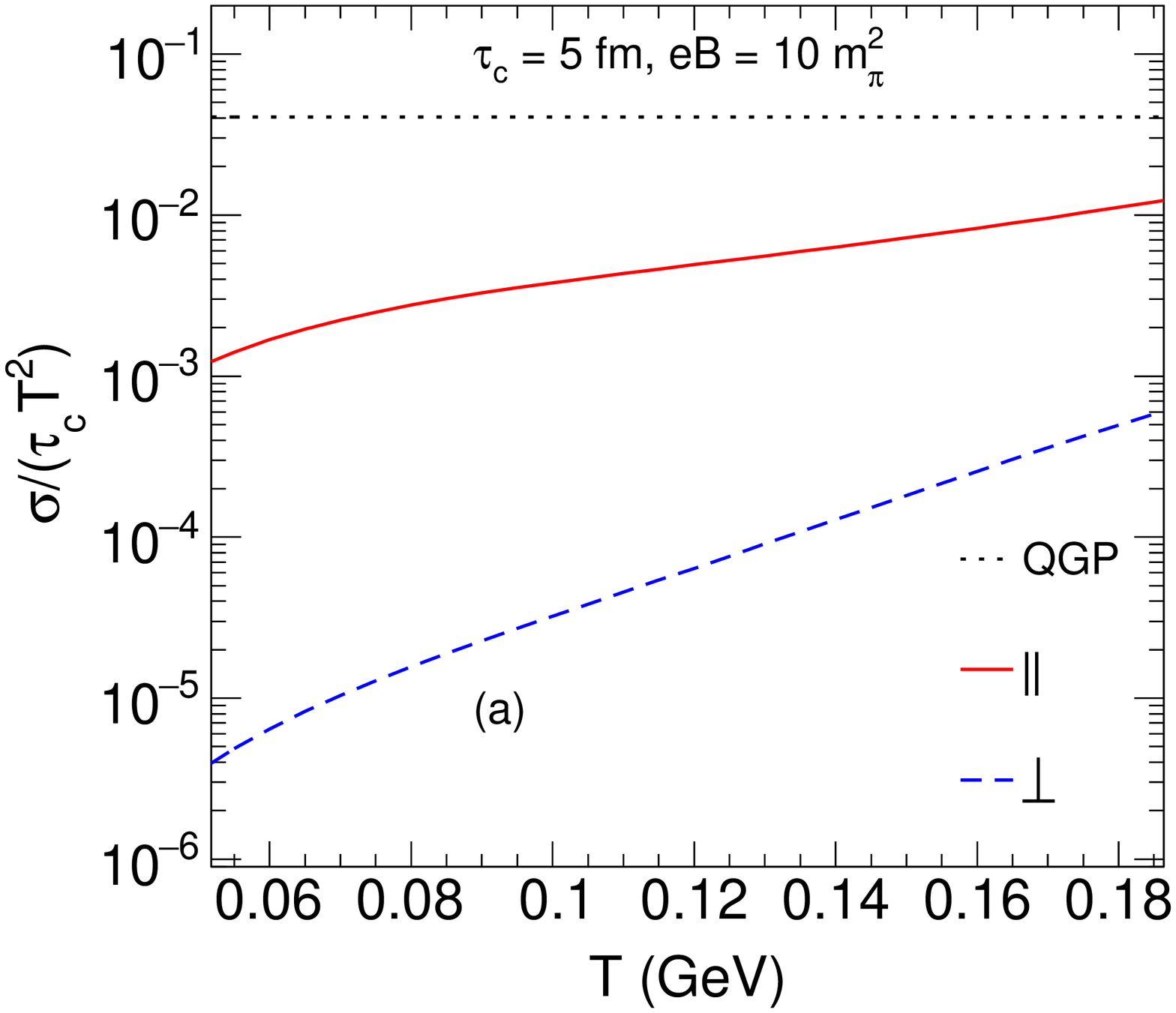}
\includegraphics[width=0.48\textwidth]{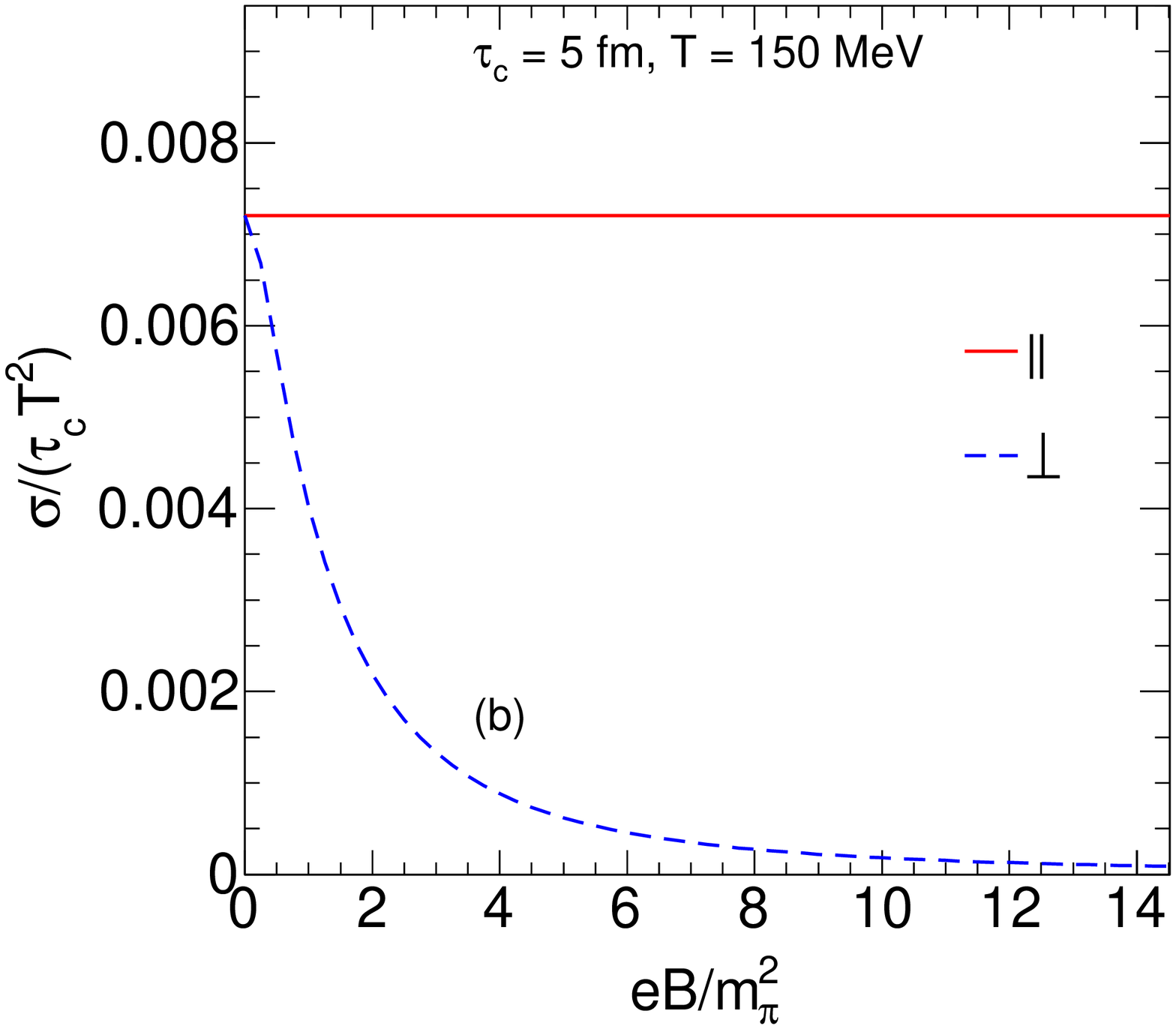}
\caption{Anisotropic component of the electrical conductivity ($\sigma_{\perp}$) for $eB =10m_{\pi}^{2}$
and its isotropic value ($\sigma_{\parallel}$) for $B=0$ are plotted as function of 
(a)temperature ($T$) and, (b) the external magnetic field ($B$).
%(c) thermal relaxation time $\tau_c$ in the framework of HRG model.
}
\label{fig:el_TB}
\end{figure*}

Next, we explore the role of $B$ and $T$ on shear viscosity as shown %by plotting $\eta/(\tau_c T^4)$ against $T$ and $B$-axes 
in Fig.~(\ref{fig:eta_TB}). For reference, we have also shown the values of $\eta/(\tau_c T^4)$ for a massless QGP (black dotted line) 
and that of HRG with $B=0$ (shown by the red solid line). 
The $\eta_{\perp}/\tau_c T^{4}$ of charged hadrons for $eB=10 m_\pi^2$ and $\tau_c=5$ fm is shown by the 
dash-dotted line in Fig.~\ref{fig:eta_TB}(a).
%
%in Fig.~\ref{fig:eta_TB}(a) for getting reference points of finite $B$ results.
%Using $\tau_c=5$ fm, $eB=10 m_\pi^2$ in Eq.~(\ref{eta_5}), we have obtained $\eta_2$
%for charged hadrons, presented by dash-dotted line in Fig.~\ref{fig:eta_TB}(a). 
%
Since HRG is composed of both charged and neutral hadrons, it is interesting to study the relative contribution 
of the charged and uncharged hadrons to the total shear viscosity.
Neutral hadrons only contribute to isotropic
shear viscosity since for neutral hadrons $\eta$ has single component, which is essentially
$\eta=\eta_{\parallel}$.
It is clear from fig.~\ref{fig:eta_TB}(a) that the anisotropic shear viscous coefficients from the charged hadrons contribution is quite smaller than
that of the isotropic shear viscosity which also contains contributions from the neutral hadrons.
However, the above fact is only true for large magnetic fields (in fig.~\ref{fig:eta_TB}(a) $B=10m_{\pi}^{2}$).
For a smaller magnetic fields the $\eta_\perp/\tau_c T^{4}$ becomes comparable or even larger than the isotropic $\eta /\tau_c T^{4}$ as shown in 
 fig.~\ref{fig:eta_TB}(b). 
 The $\parallel$ (red solid line) and $\perp$ (blue dash-double-dotted line) 
 components of shear viscosity are plotted against $B$-axis in fig.~\ref{fig:eta_TB}(b).
 The neutral hadrons contribution, which is independent of $B$ is shown by dashed line, while 
 the charged hadrons contribution is shown by dashed-dotted line. Blue dashed-double-dotted line
 is basically summation of dash (neutral hadrons) and dashed-dotted (charge hadrons) lines.
 To get some numerical estimate we note that  for $B=0$ the charged hadron contribution in
the viscosity is more than $50\%$ than the neutral hadrons.  As $B$ increases, the charge hadron
contribution decreases and for  $eB\geq 10 m_\pi^2$, 
this contribution reduces to $\sim 4\%-8\%$.
%There are large number of neutral hadrons, whose contribution (based on Eq.~\ref{Tr_B0}) is shown by dash lines.
%Since neutral hadrons follow $\propto \tau_c$ relation, whereas charge hadron follow 
%$\propto \tau_c/[1+(\tau_c/\tau_B)^2]$ relation. Due to the anisotropic factor 
%$A_{\perp}=1/[1+(\tau_c/\tau_B)^2]$, charge hadrons contribution in shear viscosity is quite suppressed than
%that of neutral hadrons and total viscosity become almost equal to viscosity of neutral hadrons.

%This fact is true for high $B$ range but not at all true for low $B$ values. This
%fact can be well understood in Fig.~\ref{fig:eta_TB}(b), showing the variation of 
%different components against $B$-axis. Neutral hadron component has field independent
%constant value, shown by dash horizontal line. At $B=0$, charged hadron contribution in
%viscosity is more than $50\%$ than neutral hadrons, but as $B$ increases, charge hadron
%contribution decreases due to the anisotropic factor $A_{\perp}$ and around high magnetic
%field range ($eB\geq 10 m_\pi^2$), more than $50\%$ contribution is shrunk to $8-4\%$.
%
\begin{figure} %{ht}
\centering
\includegraphics[width=0.48\textwidth]{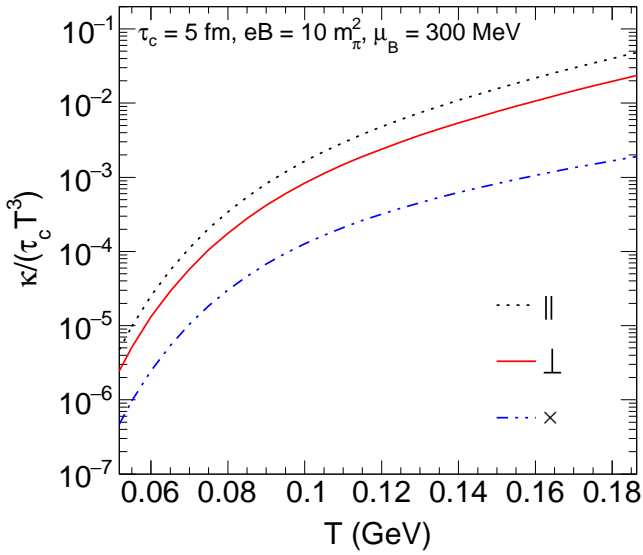}
\caption{Temperature dependence of the diffusion coefficients $\kappa_{\parallel,\perp,\times}$ 
in presence of the magnetic field.}
%and (b) shear viscosity to entropy ratio as a function of collisional relaxation time $\tau_c$.}
\label{fig:D_TB}
\end{figure}
\begin{figure} %{ht}
\centering
\includegraphics[width=0.48\textwidth]{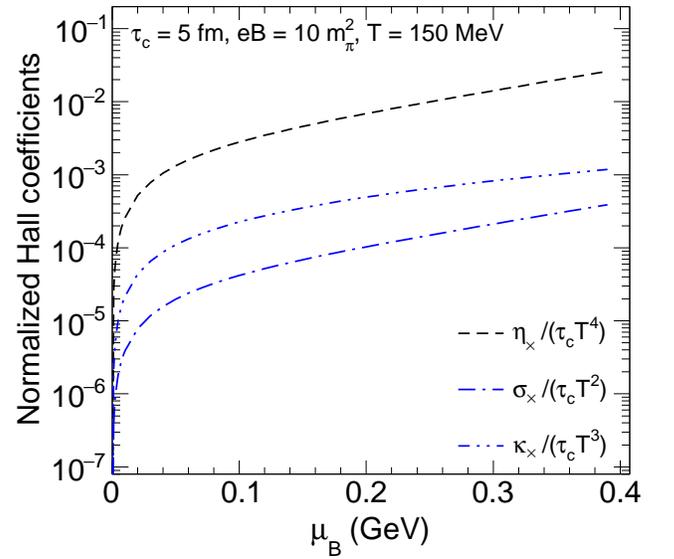}
\caption{Baryon chemical potential ($\mu_B$) dependence of (normalized) 
Hall viscosity ($\eta_{\times}$) (black dashed line), 
conductivity ($\sigma_{\times}$)(blue dashed dotted line), and the diffusion coefficients $\kappa_{\times}$ (blue dashed double dotted line) .}
\label{fig:e_s_mu}
\end{figure}
 
Let us now consider the electrical conductivity, where gluons in the QGP phase and 
the neutral hadrons in the HRG phase plays no role due to the charge neutrality. 
The results for the electrical conductivity as function of $T$ and $B$ are plotted in Figs.~\ref{fig:el_TB}(a) and (b). 
For comparison, here also we show the massless SB limit for QGP (horizontal black dotted line) and HRG (red solid line) for $B=0$.
We found that the $T$ and $B$ dependence of the electrical conductivity and the shear viscosity are very similar in nature. They mostly differ
due to the different contribution from the neutral hadron's. For example, the neutral hadrons does not contribute  to the electrical conductivity but
plays a role in the transport phenomenon related to the shear viscosity. At this point we would like to add a few 
comments: (i) we note that both $\eta/(\tau_c T^4)$, and  $\sigma/(\tau_c T^2)$ has the largest value for massless 
QGP, (ii) in presence of the magnetic field the transport coefficient becomes anisotropic and among the various components the $\parallel$ component is the largest and equals to the corresponding isotropic value of the transport coefficient (i.e., for $B=0$). (iii) there is a small difference in the temperature dependence of the isotropic and the anisotropic transport coefficients.

%{\color{red} 
%From their $T$ and $B$ dependent curves, we can draw a ranking sketch about
%their (normalized) phase-space part ($\eta/(\tau_c T^4)$, $\sigma/(\tau_c T^2)$):
%\\
%Massless QGP$(B=0)>$ HRG$(B=0)>$ HRG$(B\neq 0)$
%\\
%$\Rightarrow$ Non-Interacting$(B=0)>$ Interacting$(B=0)>$ Interacting$(B\neq 0)$~.
%\\
%Here we assume HRG as interacting QGP description, since it can well explained LQCD
%thermodynamics within hadronic temperature range. Hence, dissipation phase-space is
%reduced when one goes from non-interacting to interacting picture and this reduction
%is increased by decreasing the $T$ and increasing $B$. The interacting$(B=0)$ results,
%shown by red solid line in Figs.~(\ref{fig:eta_TB}), (\ref{fig:el_TB}) will be exactly
%equal to parallel component of dissipation coefficients like $\eta_0$, $\sigma_{\parallel}$
%in $B\neq 0$ picture. With respect to parallel component, perpendicular components $\eta_2$,
%$\sigma_{\perp}$ are smaller, which reflect anisotropic nature of medium, produced because
%of magnetic field. Ratio between perpendicular and parallel components might be considered
%as anisotropic measurement of the system, which will always be less than equal to one.
%From Figs.~(\ref{fig:eta_TB}), (\ref{fig:el_TB}), we notice that this anisotropy is
%increasing strongly as we increase $B$. It is also increased by lowering temperature but very mild way.
%}

Finally we discuss the diffusion coefficient $\kappa$.  
% Since $\kappa$ is divergent at $\mu=0$ because of its connection with 
% the enthalpy per particle $h$, therefore, we have to estimate it at non-zero $\mu$.
Similar to the electrical conductivity, in presence of a magnetic field the thermal diffusion coefficient also have
three components - $\kappa_{\parallel}$, $\kappa_{\perp}$ and $\kappa_{\times}$. 
As usual the  $\kappa_{\parallel}$ by construction is independent of the magnetic field but
$\kappa_{\perp}$ and $\kappa_{\times}$ are function of the magnetic field. 
In fig.(\ref{fig:D_TB}) we show the diffusion coefficients as a function of temperature for $B=10m_{\pi}^{2}$
and $\mu_{B}=300$. From fig.(\ref{fig:D_TB}) we see that 
$\kappa_{\perp}$ and $\kappa_{\times}$ are always smaller than $\kappa_{\parallel}$ for the temperature range 
considered here.  A non-zero Hall diffusion coefficient $\kappa_{\times}$ can be attributed to the non-zero $\mu_B$,
because for finite  $\mu_B$ the particles and the anti-particles flow due to the Hall effect do not cancel out. 
Similarly, one can get non-zero Hall shear viscosities $\eta_{\times}$, $\eta'_{\times}$ and the Hall electrical conductivity 
 for non-vanishing $\mu_B$. All of these Hall like transport coefficients vanishes for a net-baryon free medium because
 the contribution from the particles and the anti-particles are exactly equal and opposite. Fig.~(\ref{fig:e_s_mu}) demonstrate this $\mu_B$ dependent Hall viscosity ($\eta_{\times}$),
 Hall conductivity ($\sigma_{\times}$), and the Hall diffusion ($\kappa_{\times}$) for $T=150$ MeV, $eB=10m_{\pi}^{2}$, and $\tau_{c}=5$ fm. 
It is clearly seen that both $\eta_{\times}$ (black dashed line), $\sigma_{\times}$ (blue dashed dotted line)
and $\kappa_{\times}$ (blue dashed double dotted line) increase monotonically from zero at $\mu_B=0$.The growing tendency can be understood from the $\mu_B$ dependent of the net baryon density of HRG system, which is roughly proportional to $\sinh(\mu/T)$ for the Maxwell-Boltzmann distribution which at high temperature fairly well describe the Fermi-Dirac or Bose-Einstein distribution function.

The present methodology is semi-classical (as we consider quantum statistical distribution
function) in nature and does not include the Landau quantization - a quantum aspects, which is visible 
in the strong magnetic field. This effect is separately addressed in Ref.~\cite{SS_HRGB_QM},
but the complete understanding is still missing and we need further theoretical research in this direction. 
%However, the classical domain of microscopic
%calculations is itself carrying its importance because of its rich anisotropic structure. 
The physics of the anisotropic dissipation of the relativistic fluid in a magnetic field is also applicable for non-relativistic fluid,
such as different condensed matter and biological systems. 
%Our classical methodological steps for the anisotropic dissipations, which is cultivating in high energy nuclear physics field, might be useful for that domain also.

\section{Summary} 
\label{sec:sum}
In high energy heavy-ion collisions, large transient magnetic fields are produced predominantly in the 
perpendicular direction to the reaction plane. This magnetic field breaks the isotropy of the system and as a result, 
the transport coefficients become anisotropic. 
%An isotropic transport properties of medium is broken by external magnetic field. 
% We have explored this anisotropic transport properties for a HRG system. 
We evaluate the anisotropic transport coefficients of the HRG and massless QGP  by using the relaxation time approximation method. 
We use a unique tensorial decomposition of the anisotropic thermodynamic forces which
reduces the computational complexity for evaluating anisotropic transport coefficients.
%We know that HRG model calculations grossly map the temperature dependent thermodynamical quantities, obtained by LQCD calculations within hadronic temperature range. Being suppressed with respect to the massless or non-interacting QGP thermodynamics, these HRG thermodynamics can be  considered as manifestation of QCD interaction. 
%Same scenario happens for different transport coefficients like
%shear viscosity, electrical conductivity, if we normalized relaxation time. Through this normalization, we can 
%get phase-space of transport coefficients, which more or less follow similar pattern of thermodynamical phase-space.
Along with the usual relaxation time, which appears in the collision kernel of the Boltzmann equation and controls the rate of 
reaching equilibrium  for systems that are initially away from the equilibrium, in magnetic fields, we have another timescale equals to the inverse of the cyclotron frequency. 
%In presence of magnetic field, a new time scale due to magnetic field, which is basically inverse of synchrotron frequency,is come into the picture along with traditional relaxation time, already exist in dissipation picture. 
The measure of anisotropy turned out to be a function of the ratio of these two time scales. 
It is not surprising that we found the anisotropy increases  with magnetic field, and due to the specific 
choice of tensorial decomposition the $\parallel$ components of the anisotropic transport coefficients turned out to be 
the same with the isotropic case (i.e., for $B=0$).  We estimate the relative contribution of electrically charged and neutral 
hadrons to the various transport coefficients using HRG model. Since the neutral hadrons are unaffected by the Lorentz
force, they do not contribute in the anisotropic transport phenomenon. We have shown that the charged hadron contribution in 
the viscosity is more than 50\% than the neutral hadrons. As B increases, the charge hadron contribution decreases and 
for $eB \geq 10 m_{\pi}^2$ this contribution reduces to ~4\%-8\%.
%Ratio between the two time scales mainly build an anisotropic factor, for which perpendicular component of transport coefficient is reduced with respect parallel component, which remain independent of magnetic field. This difference between parallel and perpendicular components of transport coefficients create the anisotropic dissipation in medium
%and this anisotropy increases as magnetic field strength increases. So, from massless QGP to HRG or non-interacting to
%interacting QCD, dissipation phase-space face first reduction. Then from zero to non-zero magnetic field,
%dissipation phase-space in perpendicular plane face second reduction. Present article has sketched this reduced
%phase-space as a function of temperature and magnetic field through HRG model calculations. 
In case of diffusion constant we need to consider a medium with  finite $\mu_{B}$, in this study we show the result for 
$\mu_{B}=300$ MeV. We also find that non-dissipative Hall like shear viscosity and conductivity increases monotonically with $\mu_{B}$ from zero at $\mu_{B}=0$.
It turned out that there are three diffusion coefficients in non-zero magnetic fields and among them
the $\parallel$ component is the largest one.  It is interesting to note that in calculating the diffusion coefficients 
we do not  explicitly take into account the electric charge of the hadrons but we observe the anisotropic diffusion 
coefficients due to the imbalance of particle and anti-particle numbers. We also sketch chemical potential
dependence of Hall transport coefficients - how they grow from their vanishing values for (net) baryon 
free matter? These anisotropic picture of dissipations might have a broad implication in other different research fields,
where relevant impositions of system might have to be considered.

%................................................................................
{\bf Acknowledgment:} 
JD and SG acknowledge to MHRD funding facility in IIT Bhilai for supporting this theoretical work. 
UG and VR  are supported by the DST INSPIRE Faculty research grant, India. AD and VR  acknowledge 
support from the department of atomic energy, Govt. of India. SS is supported from Polish National 
Agency for Academic Exchange through Ulam Scholarship with AGREEMENT NO: PPN/ULM/2019/1/00093/U/00001. 

%%%%%%%%%%%%%%%%%%%%%%%%%%%%%%%%%%%%%%%%%%%%%%%%%%%%%%%%%%%%%%%%

%%%%%%%%%%%%%%%%% Appendix %%%%%%%%%%%%%%%%%%%%%%%%%%%%%%%%%%%%

\appendix

%%%%%%%% Subsection Electrical conductivity %%%%%%%%%%%%%%%
\section{Electrical conductivity in presence of magnetic field}
\label{sec:AppB}%
Electrical conductivity in absence of the magnetic field for a quasi-particle system having degeneracy $g$,  
electric charge $q$, four momentum $p^\mu \equiv (p^{0}, \vp)$ is~\cite{Sedrakian_el,Landau,SG_PRD}  
\bea
\sigma&=&gq^2 \frac{\beta}{3}\int \frac{d^3{\vec p}}{(2\pi)^3}\frac{p^2}{(p^{0})^2}
\tau_c f_0(1-r f_0)~,
\label{sig_B0}
\eea
where $r=\pm$ stand for the fermion/boson, $\tau_{c}$ is the thermal relaxation time.
In this section we are dealing with only one hadron species.  

For deriving the expression of the electrical conductivity in presence of a magnetic field, let's start with the Ohm's law,
\be
J^i=\sigma^{ij}E_j~,
\label{Ohm}
\ee
Here, $J^i=J^i_0 + J^i_D$, with $J^i_0$, $J^i_D$ are the ideal and the dissipative part of the three electric current density 
respectively. $\sigma^{ij}$ is the electrical conductivity tensor, $E^j$'s are the electric field components in the $j$-th
direction and $i,j$ runs from $1$ to $3$.

Now, the dissipative part of the current density according to the microscopic definition can be expressed as,
\bea
J^i_D = gq\int \frac{d^3{\vec p}}{(2\pi)^3}\frac{p^i}{p^{0}}\delta f~,
\eea
Here, $\delta f$ is deviation of the distribution function $f$ from its equilibrium part 
$f_0 = \frac{1}{e^{\beta (p^{0}-\mu)} + r}$.

Comparing the Ohm's law and the microscopic definition of the dissipative current density we get  , 
\bea
\sigma^{ij} E_j = J^i_D = gq\int \frac{d^3{\vec p}}{(2\pi)^3}\frac{p^i}{p^{0}}\delta f~.
\label{sigma}
\eea

To find the $\delta f$ we use relativistic Boltzmann equation (RBE)~\cite{Landau,Sedrakian_el,JD1,JD2},
% \be
% p^{\mu}\partial_{\mu}f + qF^{\mu \nu}p_\nu \frac{\partial f}{\partial p^\mu} = -\frac{U\cdot p}{\tau_{c}}\delta f~,
% \label{RBE_el}
% \ee
% 
\be
\frac{\partial f}{\partial t} +\frac{p^j}{p^{0}}\frac{\partial f}{\partial x^j} 
+ \frac{dp_j}{dt}\frac{\partial f}{\partial p^j} = I[\delta f]~.
\label{Boltz}
\ee

Where, $I[\delta f]$ is the linearized collision integral. Use of the relaxation time approximation 
(RTA) corresponds to  $I[\delta f] = -\frac{\delta f}{\tau_c}$ and we also note that the term  $\frac{dp_j}{dt}$ on the L.H.S of the above equation represents the force due to the electric $\vec{E}$ 
and the magnetic field $\vec{B}$. So, Eq.~(\ref{Boltz}) can be written as (assuming vanishing $\frac{\partial f}{\partial t} $ and $\frac{\partial f}{\partial x^j}$ )
\bea
\nn -q (\vec{E} +\frac{\vp}{p^{0}}\times \vec{B})\frac{\partial f}{\partial \vp} &=& -\frac{\delta f}{\tau_c}, \\
\nn \Rightarrow q\vec{E}\frac{\partial f}{\partial \vp} + (\frac{\vp}{p^{0}}\times \vec{B})\frac{\partial f}{\partial \vp} &=& \frac{\delta f}{\tau_c}, \\
\Rightarrow q\vec{E} \frac{\vp}{p^{0}} \frac{\partial f_0}{\partial p^{0}} + (\frac{\vp}{p^{0}}\times \vec{B})
\frac{\partial {(\delta f)}}{\partial \vp} &=& \frac{\delta f}{\tau_c}~.
\label{BoltzB}
%\nn -q ({\boldsymbol{E}} +\frac{\bp}{\omega}\times {\boldsymbol{B}})\frac{\partial f}{\partial \bp} &=& -\frac{\delta f}{\tau_c}\\
%
%\nn \Rightarrow q\boldsymbol{E}\frac{\partial f}{\partial \bp} + (\frac{\bp}{\omega}\times {\boldsymbol{B}})\frac{\partial f}{\partial \bp} &=& \frac{\delta f}{\tau_c}\\
%
%\Rightarrow q\boldsymbol{E} \frac{\bp}{\omega} \frac{\partial f_0}{\partial \omega} + (\frac{\bp}{\omega}\times {\boldsymbol{B}})
%\frac{\partial {(\delta f)}}{\partial \bp} &=& \frac{\delta f}{\tau_c}~.
%\label{BoltzB}
\eea
Since the second term of L.H.S., $(\frac{\vp}{p^{0}}\times \vec{B}) \frac{\vp}{p^{0}} \frac{\partial f_0}{\partial p^{0}} = 0$, %$(\frac{\bp}{\omega}\times {\boldsymbol{B}}) \frac{\bp}{\omega} \frac{\partial f_0}{\partial \omega} = 0$,
so we have considered the $\delta f$ term.

Now, we assume $\delta f=-\phi  \frac{\partial f_0}{\partial p^{0}} $, 
where $\phi = \vp \cdot \vec{F}$ with $\vec{F}=(l {\hat e} + m{\hat b} + n({\hat e}\times{\hat b})$, 
where $\hat e$ and $\hat b$ are unit vector along $\vec{E}$ and $\vec{B}$

So, Eq.~(\ref{BoltzB}) becomes
\bea
\nn \frac{1}{p^{0}}\left[
-qE \hat{e} + qB \hat{b}\times (l {\hat e} + m{\hat b} + n({\hat e}\times{\hat b}))\right] = (l {\hat e}\\
 + m{\hat b} + n({\hat e}\times{\hat b}))/\tau_c &&
\eea

%\bea
%\nn \frac{1}{\omega}\left[
%-qE \hat{e} + qB \hat{b}\times (l {\hat e} + m{\hat b} + n({\hat e}\times{\hat b}))\right] = (l {\hat e}\\
% + m{\hat b} + n({\hat e}\times{\hat b}))/\tau_c &&
%\eea
%
Now, comparing coefficients of $\hat{e}$, $\hat{b}$ and $({\hat e}\times{\hat b})$ and solving for $l$, $m$ and $n$ we get
\bea
l &=&\left(\frac{-qE\tau_c}{p^{0}}\right)\frac{1}{1+(\tau_c/\tau_B)^2}
\nn\\
m&=&\left(\frac{-qE\tau_c}{p^{0}}\right)\frac{(\tau_c/\tau_B)^2}{1+(\tau_c/\tau_B)^2}({\hat e}\cdot{\hat b})
\nn\\
n&=&\left(\frac{-qE\tau_c}{p^{0}}\right)\frac{(\tau_c/\tau_B)}{1+(\tau_c/\tau_B)^2}~,
\label{phi_coeff}
\eea
%\bea
%l &=&\left(\frac{-qE\tau_c}{\om}\right)\frac{1}{1+(\tau_c/\tau_B)^2}
%\nn\\
%m&=&\left(\frac{-qE\tau_c}{\om}\right)\frac{(\tau_c/\tau_B)^2}{1+(\tau_c/\tau_B)^2}({\hat e}\cdot{\hat b})
%\nn\\
%n&=&\left(\frac{-qE\tau_c}{\om}\right)\frac{(\tau_c/\tau_B)}{1+(\tau_c/\tau_B)^2}~,
%\label{phi_coeff}
%\eea
where $\tau_B={p_0}/(eB)$ is inverse of cyclotron frequency.

Hence, $\phi$ can be express as,
\bea
\nn \phi=\frac{q\tau_c}{1+(\tau_c/\tau_B)^2}\frac{p_i}{p^{0}}\{\delta_{ij}
-(\tau_c/\tau_B)\ep_{ijk}h_k \\
+(\tau_c/\tau_B)^2b_ib_j\}E_j~,
\eea
%\bea
%\nn \phi=\frac{e\tau_c}{1+(\tau_c/\tau_B)^2}\frac{p_i}{\om}\{\delta_{ij}
%-(\tau_c/\tau_B)\ep_{ijk}h_k \\
%+(\tau_c/\tau_B)^2b_ib_j\}E_j~,
%\eea
and,
\bea\label{Eq:deltafCond}
 \nn \delta f &=& -\phi  \frac{\partial f_0}{\partial p^{0}} = \phi \beta f_0(1-f_0)
  \nn\\
\Rightarrow \delta f &=& \frac{q\tau_c}{1+(\tau_c/\tau_B)^2}\frac{p_i}{p^{0}}\{\delta_{ij}
 -(\tau_c/\tau_B)\ep_{ijk}b_k
 \nn\\
 &+&(\tau_c/\tau_B)^2b_ib_j\}E_j \beta f_0(1-f_0)~~~~~~~~~
\eea

Now, using the above expression of $\delta f$ in Eq.~(\ref{sigma}), we get
\bea\label{Eq:deltafCond2}
\nn \sigma^{ij} &=& gq^2\beta\int \frac{d^3{\vec p}}{(2\pi)^3}\frac{p^i p^j}{(p^{0})^2}\frac{\tau_c}{1+(\tau_c/\tau_B)^2}\{\delta_{ij}
\nn\\
&-&(\tau_c/\tau_B)\ep_{ijk}b_k
+(\tau_c/\tau_B)^2b_ib_j\} f_0(1-f_0)
\nn\\
&=&\delta_{ij}\sigma_0
-\ep_{ijk}b_k\sigma_1+b_ib_j\sigma_2~,
\eea
where, 
\bea
\sigma_n&=&gq^2 \frac{\beta}{3}\int \frac{d^3{\vec p}}{(2\pi)^3}\frac{|\vp|^2}{(p^{0})^2}
\frac{\tau_c(\tau_c/\tau_B)^n}{1+(\tau_c/\tau_B)^2} f_0(1-f_0)\nn\\\
\label{sig_B}
\eea
and $n=0,1,2$. One can identify $\parallel$, $\perp$ and $\times$ components from $\sigma^n$
by using relations~\cite{Landau,Sedrakian_el,JD1,JD2}
\bea
\sigma_\parallel&=&\sigma_0 +\sigma_2 = gq^2 \frac{\beta}{3}\int \frac{d^3{\vec p}}{(2\pi)^3}\frac{|\vp|^2}{(p^{0})^2}\tau_c f_0(1-rf_0)
\nn\\
\sigma_\perp&=&\sigma_0 = gq^2 \frac{\beta}{3}\int \frac{d^3{\vec p}}{(2\pi)^3}\frac{|\vp|^2}{(p^{0})^2}
\frac{\tau_c}{1+(\tau_c/\tau_B)^2} f_0(1-rf_0)
\nn\\
\sigma_{\times}&=&\sigma_1 = gq^2 \frac{\beta}{3}\int \frac{d^3{\vec p}}{(2\pi)^3}\frac{|\vp|^2}{(p^{0})^2}
\frac{(\tau_c^2/\tau_B)}{1+(\tau_c/\tau_B)^2} f_0(1-rf_0)~.
\nn\\
\label{sig_3}
\eea
%%%%%%%%%%%%%%%%%%%%%%%%%%%%%%%%%%%%%%%%%%%%%%%%%%%
%%%%%%%%%%%%%%%%%   L.H.S  B Eqn  %%%%%%%%%%%%%%%%%
\section{Structure of RBE in RTA}\label{sec:App}
In presence of magnetic field RBE with RTA can be written as~\cite{Landau,Asutosh,JD1,JD2},
\be
p^{\mu}\partial_{\mu}f + qF^{\mu \nu}p_\nu \frac{\partial f}{\partial p^\mu} = -\frac{U\cdot p}{\tau_{c}}\delta f~,
\label{RBE}
\ee
where, $F^{\mu \nu}$ is field strength tensor, 
%and for comparetively very weak electric field 
carry only magnetic field term -
$F^{\mu \nu} =-Bb^{\mu \nu}$ with $B^{\mu\nu}=\epsilon^{\mu\nu\rho\alpha}B_\rho U_{\alpha}$. $B$ is the magnetic 
field strength and $b^\mu$ is the unit four vector. So, for a small deviation of the distribution function 
from the equilibrium, eq.~(\ref{RBE}) can be written as,
\be
p^{\mu}\partial_{\mu}f_0 = \Big(-\frac{U\cdot p}{\tau_{c}}\Big)\Big[1-\frac{qB \tau_{c}}{U\cdot p}b^{\mu\nu}p_{\nu}
\frac{\partial}{\partial p^{\mu}}\Big]\delta f~.
\label{RBET}
\ee
Equilibrium distribution function is $f_0=\frac{1}{e^{\beta(U\cdot p - \mu)}+r}$ where, chemical potential $\mu$ have space time dependency.

\begin{widetext}
So, the left hand side of the above equation can be written as, 
\bea 
p^{\mu}\partial_{\mu}f_0&=& p^{\mu}U_{\mu}Df_0+p^{\mu}\nabla_{\mu}f_0\nonumber\\
&=& \frac{\partial f_0}{\partial T}\Big(\big(U\cdot p\big)DT+p^{\mu}\nabla_{\mu}T\Big)
+\frac{\partial f_0}{\partial (\mu/T)}\Big(\big(U\cdot p\big)D\big(\frac{\mu}{T}\big)
+p^{\mu}\nabla_{\mu}\big(\frac{\mu}{T}\big)\Big)+\frac{\partial f_0}{\partial U^{\nu}}\Big(\big(U\cdot p\big)DU^{\nu}
+p^{\mu}\nabla_{\mu}U^{\nu}\Big)
\nn\\
\label{LHS}
\eea 
Where, $U^\mu$ is four velocity of particle, $D\equiv U^\mu \partial_\mu$, $\nabla^\mu\equiv \Delta^{\mu\nu}\partial_\mu$ 
with $\Delta^{\mu\nu}=g^{\mu\nu}-U^\mu U^\nu$, $g^{\mu\nu}\equiv {\rm diag}(1, -1, -1, -1)$.
Now, using the energy-momentum conservation ($\partial_\mu T_0^{\mu\nu}=0$), current conservation ($\partial_\mu N_0^\mu =0$) 
equations and the Gibbs Duhem relation we get,
\be
p^{\mu}\partial_{\mu}f_0=\frac{f_0(1-rf_0)}{T}\bigg\{Q\nabla_{\sigma}U^{\sigma}-p^{\mu}p^{\nu}[\nabla_{\mu}U_{\nu}
-\frac{1}{3}\Delta_{\mu\nu}\nabla_{\sigma}U^{\sigma}]
+\Big[1-\frac{(U\cdot p)}{h}\Big]p^{\mu}T\nabla_{\mu}\Big(\frac{\mu}{T}\Big)\bigg\}
\ee
Where $Q=(U\cdot p)^{2} (\frac{4}{3}-\gamma^{'})+(U\cdot p)\Big[(\gamma^{''}-1)h-\gamma^{'''}T\Big]-\frac{1}{3}m^{2}$ 
and $h=m S_{3}^{1} / S_{2}^{1}$. The expressions for $\gamma^{'}$, $\gamma^{''}$ , $\gamma^{'''}$ and $S_{n}^{\alpha}$ are 
\bea
\gamma^{\prime}&=&\frac{\left(S_{2}^{0} / S_{2}^{1}\right)^{2}-\left(S_{3}^{0} / S_{2}^{1}\right)^{2}+4 z^{-1} S_{2}^{0} 
S_{3}^{1} /\left(S_{2}^{1}\right)^{2}+z^{-1} S_{3}^{0} / S_{2}^{1}}{\left(S_{2}^{0} / S_{2}^{1}\right)^{2}-\left(S_{3}^{0} / 
S_{2}^{1}\right)^{2}+3 z^{-1} S_{2}^{0} S_{3}^{1} /\left(S_{2}^{1}\right)^{2}+2 z^{-1} S_{3}^{0} / S_{2}^{1}-z^{-2}}
\nn\\
\gamma^{\prime \prime}&=&1+\frac{z^{-2}}{\left(S_{2}^{0} / S_{2}^{1}\right)^{2}-\left(S_{3}^{0} / S_{2}^{1}\right)^{2}
+3 z^{-1} S_{2}^{0} S_{3}^{1} /\left(S_{2}^{1}\right)^{2}+2 z^{-1} S_{3}^{0} / S_{2}^{1}-z^{-2}}
\nn\\
\gamma^{\prime \prime \prime}&=&\frac{S_{2}^{0} / S_{2}^{1}+5 z^{-1} S_{3}^{1} / S_{2}^{1}-S_{3}^{0} S_{3}^{1} 
/\left(S_{2}^{1}\right)^{2}}{\left(S_{2}^{0} / 
S_{2}^{1}\right)^{2}-\left(S_{3}^{0} / S_{2}^{1}\right)^{2}+3 z^{-1} S_{2}^{0} S_{3}^{1} 
/\left(S_{2}^{1}\right)^{2}+2 z^{-1} S_{3}^{0} / S_{2}^{1}-z^{-2}}
\eea
where $z=m/T$ and $S_{n}^{\alpha}(z)=\sum_{k=1}^{\infty}(-r)^{k-1} e^{k \mu / T} k^{-\alpha} K_{n}(k z)$, $K_{n}(x)$ 
denoting the modified Bessel function of order $n$.
\end{widetext}
% 
% 
% Where,
% $Q=(U\cdot p)^{2} (\frac{4}{3}-\gamma^{'})+(U\cdot p)\Big[(\gamma^{''}-1)h-\gamma^{'''}T\Big]-\frac{1}{3}m^{2}$ and $h=m S_{3}^{1} / S_{2}^{1}$ {\color{red}enthalpy per particle}? 
% The expressions for $\gamma^{'}$, $\gamma^{''}$ , $\gamma^{'''}$ and $S_{n}^{\alpha}$ are {\color{red}given in Appendix-A??}
%%%%%%%%%%%%%%%%%%%%%%%%%%%%%%%%%%%%%%%%%%%%%%%%%%%

%%%%%%%%%%%%%%%%% Shear Viscosity %%%%%%%%%%%%%%%%%
\subsection{Shear Viscosity}
In presence of magnetic field, the general expression of $\delta f$ for shear viscosity is considered as~
\bea\label{Eq:deltafShear}
\delta f&=&\sum\limits_{n=0}^{4} c_{n}C_{(n)\mu\nu\alpha\beta}p^{\mu}p^{\nu}V^{\alpha\beta}\\
&=&\Big[ c_{0} P^{0}_{\langle \mu\nu\rangle\alpha\beta}+ c_{1}\big(P^{1}_{\langle \mu\nu\rangle\alpha\beta}+P^{-1}_{\langle \mu\nu\rangle\alpha\beta}\big) \nonumber\\
&+& ic_{2}\big(P^{1}_{\langle \mu\nu\rangle\alpha\beta}-P^{-1}_{\langle \mu\nu\rangle\alpha\beta}\big) + c_{3}\big(P^{2}_{\langle \mu\nu\rangle\alpha\beta}+P^{-2}_{\langle \mu\nu\rangle\alpha\beta}\big)\nonumber\\
&+& ic_{4}\big(P^{2}_{\langle \mu\nu\rangle\alpha\beta}-P^{-2}_{\langle \mu\nu\rangle\alpha\beta}\big) \Big]p^{\mu}p^{\nu}V^{\alpha\beta}~,
\eea
where $V_{\alpha \beta} = \frac{1}{2}(\frac{\partial U_{\alpha}}{\partial x_{\beta}} 
+ \frac{\partial U_{\beta}}{\partial x_{\alpha}})$, and $P^{(\mathrm{m})}_{\langle \mu\nu\rangle\alpha\beta}=P^{(\mathrm{m})}_{\mu\nu\alpha\beta}+P^{(\mathrm{m})}_{\nu\mu\alpha\beta}$.
The fourth rank projection tensor is defined in terms of the second rank projection tensor as \cite{Hess},
\bea
P_{\mu \nu, \mu^{\prime} \nu^{\prime}}^{(\mathrm{m})}=\sum_{\mathrm{m}_{1}=-1}^{1} \sum_{\mathrm{m}_{2}=-1}^{1} P_{\mu \mu^{\prime}}^{\left(\mathrm{m}_{1}\right)} P_{\nu \nu^{\prime}}^{\left(\mathrm{m}_{2}\right)} \delta\left(m, m_{1}+m_{2}\right).
\eea 
and the second rank projection tensor is defined as,
\bea
P^{0}_{\mu\nu}&=&b_{\mu}b_{\nu} ,
\nn\\ 
P^{1}_{\mu\nu}&=&\frac{1}{2} \left( \Delta_{\mu\nu}- b_{\mu}b_{\nu}+ib_{\mu\nu} \right) ,
\nn\\
P^{-1}_{\mu\nu}&=&\frac{1}{2} \left( \Delta_{\mu\nu}- b_{\mu}b_{\nu}-ib_{\mu\nu} \right).
\nn
\eea 
where $\Delta^{\mu\nu}=g^{\mu\nu}-u^{\mu}u^\nu$. The second rank projection tensor satisfies the following properties,
\bea
P_{\mu \kappa}^{(\mathrm{m})} P_{\kappa \nu}^{(\mathrm{m}^{\prime})} &=&\delta_{\mathrm{mm}^{\prime}} P_{\mu \nu}^{(\mathrm{m})}, \label{Prod_rule}\\
\left(P_{\mu \nu}^{(\mathrm{m})}\right)^{\dagger} &=&P_{\mu \nu}^{(-\mathrm{m})}=P_{\nu \mu}^{(\mathrm{m})}, \\
\sum_{\mathrm{m}=-1}^{1} P_{\mu \nu}^{(\mathrm{m})} &=&\delta_{\mu \nu}, \quad P_{\mu \mu}^{(\mathrm{m})}=1.
\eea
Substituting the above expression on the right hand side of the Boltzmann transport equation (\ref{RBET}) we get,
\begin{widetext}
 \bea
\Big(-\frac{U\cdot p}{\tau_{C}}\Big)\Big[1-\frac{qB \tau_{C}}{U\cdot p}b^{\mu\nu}p_{\nu}\frac{\partial}{\partial p^{\mu}}\Big]\delta f &=&\Big(-\frac{U\cdot p}{\tau_{C}}\Big)\Big[1-\frac{qB \tau_{C}}{U\cdot p}b^{\mu\nu}p_{\nu}\frac{\partial}{\partial p^{\mu}}\Big]\sum\limits_{n=0}^{4} c_{n}C_{(n)\alpha\beta\rho\sigma}p^{\alpha}p^{\beta}V^{\rho\sigma}\nonumber\\
&=&\Big(-\frac{U\cdot p}{\tau_{C}}\Big)\Big[p^{\alpha}p^{\beta}V^{\rho\sigma}\sum\limits_{n=0}^{4} c_{n}C_{(n)\alpha\beta\rho\sigma}\\
 &-&\frac{qB \tau_{C}}{U\cdot p}b^{\mu\nu}p_{\nu}\big(\Delta^{\alpha}_{\mu}p^{\beta}+\Delta^{\beta}_{\mu}p^{\alpha}\big)V^{\rho\sigma}\sum\limits_{n=0}^{4} c_{n}C_{(n)\alpha\beta\rho\sigma}\Big]=T_{1}+T_{2}
\eea
%\end{widetext}
where,
\bea
T_{1}=&\Big(-\frac{U\cdot p}{\tau_{C}}\Big)\Big[p^{\alpha}p^{\beta}V^{\rho\sigma}\sum\limits_{n=0}^{4} c_{n}C_{(n)\alpha\beta\rho\sigma}\Big] \:\:
\mathrm{and},\:
T_{2}=& qB b^{\mu\nu}p_{\nu}\big(\Delta^{\alpha}_{\mu}p^{\beta}+\Delta^{\beta}_{\mu}p^{\alpha}\big)V^{\rho\sigma}\sum\limits_{n=0}^{4} c_{n}C_{(n)\alpha\beta\rho\sigma}.
\eea
Now, 
\bea
T_{1}=& \Big(-\frac{U\cdot p}{\tau_{C}}\Big)p^{\alpha}p^{\beta}V^{\rho\sigma}\Big[ c_{0} P^{0}_{\langle\alpha\beta\rangle\rho\sigma}+ c_{1}\big(P^{1}_{\langle\alpha\beta\rangle\rho\sigma}+P^{-1}_{\langle\alpha\beta\rangle\rho\sigma}\big)
+ic_{2}\big(P^{1}_{\langle\alpha\beta\rangle\rho\sigma}-P^{-1}_{\langle\alpha\beta\rangle\rho\sigma}\big) + c_{3}\big(P^{2}_{\langle\alpha\beta\rangle\rho\sigma}+P^{-2}_{\langle\alpha\beta\rangle\rho\sigma}\big)\nonumber\\
&+ ic_{4}\big(P^{2}_{\langle\alpha\beta\rangle\rho\sigma}-P^{-2}_{\langle\alpha\beta\rangle\rho\sigma}\big) \Big].
\eea
%\begin{widetext}
 and,
\bea
T_{2}= qB b^{\mu\nu}p_{\nu}\big(\Delta^{\alpha}_{\mu}p^{\beta}+\Delta^{\beta}_{\mu}p^{\alpha}\big)V^{\rho\sigma}\sum\limits_{n=0}^{4} c_{n}C_{(n)\alpha\beta\rho\sigma}=2qB b^{\mu\nu}p_{\nu}\Delta^{\alpha}_{\mu}p^{\beta}V^{\rho\sigma}\sum\limits_{n=0}^{4} c_{n}C_{(n)\alpha\beta\rho\sigma}
\eea
Since, $C_{(n)\alpha\beta\rho\sigma}=C_{(n)\beta\alpha\rho\sigma}$. \\
So,
\bea
T_{2}&=&2qB b^{\mu\nu}p_{\nu}\Delta^{\alpha}_{\mu}p^{\beta}V^{\rho\sigma}\Big[ c_{0} P^{0}_{\langle\alpha\beta\rangle\rho\sigma}+ c_{1}\big(P^{1}_{\langle\alpha\beta\rangle\rho\sigma}+P^{-1}_{\langle\alpha\beta\rangle\rho\sigma}\big)+ ic_{2}\big(P^{1}_{\langle\alpha\beta\rangle\rho\sigma}-P^{-1}_{\langle\alpha\beta\rangle\rho\sigma}\big) + c_{3}\big(P^{2}_{\langle\alpha\beta\rangle\rho\sigma}+P^{-2}_{\langle\alpha\beta\rangle\rho\sigma}\big)\nonumber\\
&+&ic_{4}\big(P^{2}_{\langle\alpha\beta\rangle\rho\sigma}-P^{-2}_{\langle\alpha\beta\rangle\rho\sigma}\big) \Big].
\label{T1} \\
T_{2}&=&2qB V^{\rho\sigma}p_{\mu}p_{\nu}\Big[i(P^{2\langle\mu\nu\rangle\alpha\beta}-P^{-2\langle\mu\nu\rangle\alpha\beta})+\frac{i}{2}(P^{1\langle\mu\nu\rangle\alpha\beta}-P^{-1\langle\mu\nu\rangle\alpha\beta})\Big]\Big[ c_{0} P^{0}_{\langle\alpha\beta\rangle\rho\sigma}+ c_{1}\big(P^{1}_{\langle\alpha\beta\rangle\rho\sigma}+P^{-1}_{\langle\alpha\beta\rangle\rho\sigma}\big) \nonumber\\
&+& ic_{2}\big(P^{1}_{\langle\alpha\beta\rangle\rho\sigma}-P^{-1}_{\langle\alpha\beta\rangle\rho\sigma}\big)+ c_{3}\big(P^{2}_{\langle\alpha\beta\rangle\rho\sigma}+P^{-2}_{\langle\alpha\beta\rangle\rho\sigma}\big)+ ic_{4}\big(P^{2}_{\langle\alpha\beta\rangle\rho\sigma}-P^{-2}_{\langle\alpha\beta\rangle\rho\sigma}\big) \Big].\nonumber\\
&=&2qB V^{\rho\sigma}p_{\mu}p_{\nu}\Big[c_{0}\cdot 0+\frac{i}{2} c_{1}(P^{1\langle\mu\nu\rangle}_{\rho\sigma}-P^{-1\langle\mu\nu\rangle}_{\rho\sigma})-\frac{1}{2}c_{2}(P^{1\langle\mu\nu\rangle}_{\rho\sigma}+P^{-1\langle\mu\nu\rangle}_{\rho\sigma})+c_{3}(P^{2\langle\mu\nu\rangle}_{\rho\sigma}-P^{-2\langle\mu\nu\rangle}_{\rho\sigma})-c_{4}(P^{2\langle\mu\nu\rangle}_{\rho\sigma}+P^{-2\langle\mu\nu\rangle}_{\rho\sigma})\nonumber\\
&=&2qB V^{\rho\sigma}p_{\mu}p_{\nu}\Big[P^{1\langle\mu\nu\rangle}_{\rho\sigma}\Big(\frac{i}{2}c_{1}-\frac{1}{2}c_{2}\Big)+P^{1\langle\mu\nu\rangle}_{\rho\sigma}\Big(-\frac{i}{2}c_{1}-\frac{1}{2}c_{2}\Big)+P^{2\langle\mu\nu\rangle}_{\rho\sigma}\big(ic_{3}-c_{4}\big)+P^{-2\langle\mu\nu\rangle}_{\rho\sigma}\big(-ic_{3}-c_{4}\big)\Big]
\label{T2}
\eea
\end{widetext}
The left hand side of the RBE equation, neglecting the terms that include the spatial gradients of temperature and chemical potential 
in terms of the projection operator $P^{n}_{\langle\mu\nu\rangle\alpha\beta}$ turns out to be,
\bea
T_1+T_2=-\frac{f_0(1-rf_0)}{T}p^{\mu}p^{\nu}V^{\rho\sigma}\Big[P^{0}_{\langle\mu\nu\rangle\alpha\beta}+P^{1}_{\langle\mu\nu\rangle\alpha\beta}\nonumber\\
+P^{-1}_{\langle\mu\nu\rangle\alpha\beta}+P^{2}_{\langle\mu\nu\rangle\alpha\beta}
+P^{-2}_{\langle\mu\nu\rangle\alpha\beta}\Big]
\nn\\
\eea
Now equating the right hand side with the left hand side of relativistic Boltzmann equation (\ref{RBET}) with the help of eq.~(\ref{T2}),(\ref{T1}) and (\ref{LHS}) we get;
\bea
c_{0}&=&\frac{1}{2}\frac{f_0(1-rf_0)\tau_{c}}{T(U\cdot p)},
\nn\\
c_{1}&=&\frac{1}{2}\frac{(U\cdot p)f_0(1-rf_0)\tau_{c}}{T[(U\cdot p)^{2}+(qB \tau_{c})^{2}]},
\nn\\
c_{2}&=&\frac{1}{2}\frac{(qB)f_0(1-rf_0)\tau_{c}^{2}}{T[(U\cdot p)^{2}+(qB \tau_{c})^{2}]},
\nn\\
c_{3}&=&\frac{1}{2}\frac{(U\cdot p)f_0(1-rf_0)\tau_{c}}{T[(U\cdot p)^{2}+(2qB \tau_{c})^{2}]},
\nn\\
c_{4}&=&\frac{(qB)f_0(1-rf_0)\tau_{c}^{2}}{T[(U\cdot p)^{2}+(2qB \tau_{c})^{2}]}.
\eea
Using the above expressions the shear viscosities turns out to be,
\bea
\eta_{\parallel}=\frac{2}{15}\int\frac{d^3{\vec p}}{(2\pi)^{3}p_{0}}|\vp|^4c_{0}, \\
\eta_{\perp}=\frac{2}{15}\int\frac{d^3{\vec p}}{(2\pi)^{3}p_{0}}|\vp|^4c_{1},\\
\eta_{\times}=\frac{2}{15}\int\frac{d^3{\vec p}}{(2\pi)^{3}p_{0}}|\vp|^4c_{2},\\
\eta'_{\perp}=\frac{2}{15}\int\frac{d^3{\vec p}}{(2\pi)^{3}p_{0}}|\vp|^4c_{3},\\
\eta'_{\times}=\frac{2}{15}\int\frac{d^3{\vec p}}{(2\pi)^{3}p_{0}}|\vp|^4c_{4}.
\eea
\iffalse
\bea
\eta_{\parallel}&=&\frac{1}{15T}\int\frac{d^3{\vec p}}{(2\pi)^{3}}\frac{|\vp|^4}{p_0^2} \tau_{c} f_0(1-rf_0) 
\nn\\
\eta_{\perp}&=&\frac{1}{15T}\int\frac{d^3{\vec p}}{(2\pi)^{3}}\frac{|\vp|^4}{p_0^2}\frac{\tau_{c}}{1+(\tau_c/\tau_B)^2} f_0(1-rf_0)
\nn\\
\eta'_{\perp}&=&\frac{1}{15T}\int\frac{d^3{\vec p}}{(2\pi)^{3}}\frac{|\vp|^4}{p_0^2} \frac{\tau_{c}}{1+(2\tau_c/\tau_B)^2} f_0(1-rf_0)
\nn\\
\eta_{\times}&=&\frac{1}{15T}\int\frac{d^3{\vec p}}{(2\pi)^{3}}\frac{|\vp|^4}{p_0^2}
\frac{\tau_c^2/\tau_B}{1+(\tau_c/\tau_B)^2} f_0(1-rf_0)
\nn\\
\eta'_{\times}&=&\frac{1}{15T}\int\frac{d^3{\vec p}}{(2\pi)^{3}}\frac{|\vp|^4}{p_0^2}\frac{\tau_c^2/\tau_B}{\frac{1}{2}
+2(\tau_c/\tau_B)^2}f_0(1-rf_0)\nn\\
\eea
\fi
%%%%%%%%%%%%%%%%%%%%%%%%%%%%%%%%%%%%%%%%%%%%%%%%%%%

%%%%%%%%%%%%%%%%% Bulk Viscosity %%%%%%%%%%%%%%%%%%
\subsection{Bulk Viscosity}
\begin{widetext}
As mentioned earlier in the text, in presence of magnetic field there is three components of the bulk viscosity and the form of $\delta f$ corresponds to them is
\bea\label{Eq:deltafBulk}
\delta f=\sum_{n=1}^{3}c_{n}C_{(n)\mu\nu}\partial^{\mu}U^{\nu}=\left(c_{1}P^{0}_{\mu\nu} +c_{2}(P^{1}_{\mu\nu}+P^{-1}_{\mu\nu})
+c_{3}(P^{1}_{\mu\nu}-P^{-1}_{\mu\nu})\right)\partial^{\mu}U^{\nu}.
\eea 
So, the right hand side of RBE becomes 
\bea
-\frac{U\cdot p}{\tau_c}\Big[1-\frac{qB \tau_{c}}{(U\cdot p)}b^{\mu\nu}p_{\nu}\frac{\partial}{\partial p^{\mu}} \Big]\delta f
&=&-\frac{U\cdot p}{\tau_c}\Big[1-\frac{qB \tau_{c}}{(U\cdot p)}b^{\mu\nu}p_{\nu}\frac{\partial}{\partial p^{\mu}} \Big]\big\{c_{1}(b^{\mu}b^{\nu}) 
+ c_{2}(\Delta^{\mu\nu}-b^{\mu}b^{\nu})+ic_{3}b^{\mu\nu}\big\}\partial_{\mu}U_{\nu} \nonumber\\
&=&-\frac{U\cdot p}{\tau_c}\big\{c_{1}(b^{\mu}b^{\nu})+c_{2}(\Delta^{\mu\nu} - b^{\mu}b^{\nu})+ic_{3}b^{\mu\nu}\big\}\partial_{\mu}U_{\nu} \nonumber\\
&=&-\frac{U\cdot p}{\tau_c}\big\{c_{2}(\partial^{\mu}U_{\mu})+(c_{1}-c_{2})b^{\mu}b^{\nu}\partial_{\mu}U_{\nu} + ic_{3}b^{\mu\nu}\partial_{\mu}U_{\nu}\big\}.
\label{RHSbulk}
\eea 
\end{widetext}

Equating the coefficients of $\partial^{\mu}U_{\mu}$, $b^{\mu}b^{\nu}\partial_{\mu}U_{\nu}$ and $b^{\mu\nu}\partial_{\mu}U_{\nu}$ from eq.~(\ref{RHSbulk}) and (\ref{LHS}) we get,
\bea
c_{1}=&\frac{\tau_{C}Q}{(U\cdot p)}\frac{f_0(1-rf_0)}{T}\\
c_{2}=&\frac{\tau_{C}Q}{(U\cdot p)}\frac{f_0(1-rf_0)}{T}\\
c_{3}=&0
\eea
Thus the bulk viscosity can be derived from the relation,
\bea
\Pi^{\mu\nu}=\Pi\Delta^{\mu\nu}=\int \frac{d^3{\vec p}}{(2\pi)^{3}p^{0}}p^{\mu}p^{\nu}\delta f
\eea
$\Pi$ is known as bulk pressure.  Therefore,
\bea
\Pi&=&\frac{1}{3}\int \frac{d^3{\vec p}}{(2\pi)^{3}p^{0}}\Delta_{\mu\nu}p^{\mu}p^{\nu}\big\{c_{1}(b^{\alpha}b^{\beta})\nonumber\\
&+&c_{2}(\Delta^{\alpha\beta}-b^{\alpha}b^{\beta})+c_{3}b^{\alpha\beta}\big\}\partial_{\alpha}U_{\beta}%\nonumber\\
%=&\frac{1}{3}\int \frac{d^3\bp}{(2\pi)^{3}p^{0}}\Delta_{\mu\nu}p^{\mu}p^{\nu}\frac{Q\tau_{C}}{(U\cdot p)}\Delta^{\alpha\beta}\partial_{\alpha}U_{\beta}.
\eea
So, there components of bulk viscosity in presence of magnetic field are
\bea
\zeta_{\parallel}=\zeta_{\perp}&=&\frac{\tau_c}{T}\int \frac{d^3{\vec p}}{(2\pi)^{3}(p^{0})^{2}}Q^{2}f_0(1-rf_0)\\
\zeta_{\times}&=&0
%-\tau_c\bigg(\frac{1}{3}\left(1-3 c_{s}^{2}\right)(\epsilon+P)-\nonumber\\
%&&\frac{2}{9}(\epsilon-3 P)-\frac{gm^{4}}{9}\int \frac{d^3{\vec p}}{{(2\pi)}^3E^3} (f_0+\bar{f_0})\bigg)
\eea
Where $Q$ is already addressed in earlier subsection. Since without magnetization, there will be no
magnetic field dependent component of bulk viscosity, so its numerical results have not been explored.
%%%%%%%%%%%%%%%%%%%%%%%%%%%%%%%%%%%%%%%%%%%%%%%%%%%

%%%%%%%%%%%%%%%%% Diffusion coefficients %%%%%%%%%%%%%%%%
\subsection{Diffusion coefficient}
In presence of magnetic field for thermal diffusion component of $\delta f$ can be written as
\bea
\delta f=K^{\mu\nu}p_{\mu}\partial_{\nu}\alpha_{0}; 
\eea 
where, \(\alpha_{0}=\frac{\mu}{T} \).

The second order tensor $K^{\mu\nu}$ can be break down into the new projectors:
\bea
P^{\parallel}_{\mu\nu}&=&P^{0}_{\mu\nu}=b_{\mu}b_{\nu}, 
\nn\\
P^{\perp}_{\mu\nu}&=&\left(P^{1}_{\mu\nu}+P^{-1}_{\mu\nu}\right)=\left(\Delta_{\mu\nu}-b_{\mu}b_{\nu}\right), 
\nn\\
 P^{\times}_{\mu\nu}&=&\left(P^{1}_{\mu\nu}-P^{-1}_{\mu\nu}\right)=ib_{\mu\nu} .
\eea

Using this projectors the $\delta f$ becomes,
\bea\label{Eq:deltafDiff}
\delta f &=& \left[K_{\parallel}P^{0}_{\mu\nu}+K_{\perp}P^{\perp}_{\mu\nu} + K_{\times}P^{\times}_{\mu\nu}\right]p^{\mu}\partial^{\nu}\alpha_{0} \nonumber \\
&=&\big[K_{\parallel}b_{\mu}b_{\nu}+K_{\perp}\left(\Delta_{\mu\nu}-b_{\mu}b_{\nu} \right) \nonumber\\
& & +K_{\times}\left(ib_{\mu\nu}\right) \big] p^{\mu}\partial^{\nu}\alpha_{0}.  
\eea
\begin{widetext}
Now, with this $\delta f$ the right hand side of the Boltzmann transport equation becomes,
\bea
-\frac{U\cdot p}{\tau_{C}}\left[1-\frac{qB \tau_{C}}{\left(U\cdot p\right)}b^{\mu\nu}p_{\nu}\frac{\partial}{\partial p^{\mu}}\right]\delta f  &=& -\frac{U\cdot p}{\tau_{C}}\left[1-\frac{qB \tau_{C}}{\left(U\cdot p\right)}b^{\mu\nu}p_{\nu}\frac{\partial}{\partial p^{\mu}}\right]\big[ K_{\parallel}b_{\alpha}b_{\beta}+K_{\perp}\left(\Delta_{\alpha\beta}-b_{\alpha}b_{\beta}\right) +K_{\times}\left(ib_{\alpha\beta}\right)\big]p^{\alpha}\partial^{\beta}\alpha_{0}  \nonumber\\
&=&-\frac{U\cdot p}{\tau_{C}}\left[p^{\alpha}-\frac{qB\tau_{C}}{\left(U\cdot p\right)}b^{\mu\nu}p_{\nu}\delta^{\alpha}_{\mu}\right]\big[K_{\parallel}b_{\alpha}b_{\beta}+K_{\perp}\left(\Delta_{\alpha\beta}-b_{\alpha}b_{\beta}\right) +K_{\times}\left(ib_{\alpha\beta}\right)\big]\partial^{\beta}\alpha_{0} \nonumber\\
&=&-\frac{U\cdot p}{\tau_{C}}\big[K_{\parallel}b_{\alpha}b_{\beta}+K_{\perp}\left(\Delta_{\alpha\beta}-b_{\alpha}b_{\beta}\right)+K_{\times}\left(ib_{\alpha\beta}\right)\big]p^{\alpha}\partial^{\beta}\alpha_{0}\nn\\
&+&qB p_{\nu}\big[b^{\alpha\nu}K_{\parallel}b_{\alpha}b_{\beta}+b^{\alpha\nu}K_{\perp}\left(\Delta_{\alpha\beta}-b_{\alpha}b_{\beta}\right) +b^{\alpha\nu}K_{\times}\left(ib_{\alpha\beta}\right)\big]p^{\alpha}\partial^{\beta}\alpha_{0}  
\label{RHSDiff}
\eea

%Now since,
%\bea
%P^{\left(m\right)}_{\mu k}P^{\left(n\right)}_{k\nu}=\delta_{mn}P^{\left(m\right)}_{\mu \nu} 
%\eea

Using relation Eq. (\ref{Prod_rule})we have,
\bea
b^{\alpha\nu}P^{\parallel}_{\alpha\beta}&=&-iP^{\times \alpha\nu}P^{\parallel}_{\alpha\beta} 
=-i\left[P^{1\alpha\nu}-P^{-1\alpha\nu}\right]P^{0}_{\alpha\beta} =0;  \\
b^{\alpha\nu}P^{\perp}_{\alpha\beta}&=&-iP^{\times \alpha\nu}P^{\perp}_{\alpha\beta}
=-i\left[P^{1\alpha\nu}-P^{-1\alpha\nu}\right]\left[P^{1}_{\alpha\beta}+P^{-1}_{\alpha\beta}\right] 
=-i\left[P^{1\nu}_{\beta}-P^{2\nu}_{\beta}\right]=-iP^{\times \nu}_{\beta};\\
b^{\alpha\nu}P^{\times}_{\alpha\beta}&=&-iP^{\times \alpha\nu}P^{\times}_{\alpha\beta}
=-i\left[P^{1\alpha\nu}-P^{-1\alpha\nu}\right]\left[P^{1}_{\alpha\beta}-P^{-1}_{\alpha\beta}\right] 
=-i\left[P^{1\nu}_{\beta}+P^{-1\nu}_{\beta}\right]=-iP^{\perp\nu}_{\beta}.   
\eea 

Using the above expressions in eq.~(\ref{RHSDiff}) R.H.S. of RBE becomes,
\bea
-\frac{U\cdot p}{\tau_{C}}\left[1-\frac{qB \tau_{C}}{\left(U\cdot p\right)}b^{\mu\nu}p_{\nu}\frac{\partial}{\partial p^{\mu}}\right]\delta f&=&-\frac{U\cdot p}{\tau_{C}}\left[K_{\parallel}P^{\parallel}_{\nu\beta}p^{\nu}+K_{\perp}P^{\perp}_{\nu\beta}p^{\nu} +K_{\times}P^{\times}_{\nu\beta}p^{\nu}\right]\partial^{\beta}\alpha_{0} \nn \\ 
 &+&qB\left[K_{\parallel}0\cdot p^{\nu}-iK_{\times}P^{\perp}_{\nu\beta}p^{\nu} -iK_{\perp}P^{\times}_{\nu\beta}p^{\nu}\right]\partial^{\beta}\alpha_{0}  \nonumber\\
&=&\partial^{\beta}\alpha_{0} \bigg\{K_{\parallel}\left( -\frac{U\cdot p}{\tau_{c}}\right)P^{\parallel}_{\nu\beta}p^{\nu}-\left[\frac{U\cdot p}{\tau_{c}}K_{\perp}+iqB K_{\times}\right]P^{\perp}_{\nu\beta}p^{\nu}
\nn \\ 
&-&\left[\frac{U\cdot p}{\tau_{c}}K_{\times}+iqB K_{\perp}\right]P^{\times}_{\nu\beta}p^{\nu} \bigg\}   \nonumber\\
&=&\partial^{\beta}\alpha_{0} \bigg\{K_{\parallel}\left( -\frac{U\cdot p}{\tau_{c}}\right)b_{\nu}b_{\beta}p^{\nu}-\big[\frac{U\cdot p}{\tau_{c}}K_{\perp}+iqB K_{\times}\big]\left(\Delta_{\nu\beta}-b_{\nu}b_{\beta}\right)p^{\nu}
\nn \\
&-&\big[\frac{U\cdot p}{\tau_{c}}K_{\times}+iqB K_{\perp}\big]i b_{\nu\beta}p^{\nu} \bigg\}  \nonumber \\
&=&\partial^{\beta}\alpha_{0}\bigg\{\left[K_{\parallel}\left( -\frac{U\cdot p}{\tau_{c}}\right)+\left(\frac{U\cdot p}{\tau_{c}}K_{\perp}+iqB K_{\times}\right)\right]b_{\nu}b_{\beta}p^{\nu}-\left[\frac{U\cdot p}{\tau_{c}}K_{\perp}+iqB K_{\times}\right]\Delta_{\nu\beta}p^{\nu} 
\nn \\
&-&\big[\frac{U\cdot p}{\tau_{c}}K_{\times}+iqB K_{\perp}\big]i b_{\nu\beta}p^{\nu} \Big\} .  
\label{RHSDiff1}
\eea

So, from eq.~(\ref{RHSDiff1}) and (\ref{LHS}) the RBE becomes,

\bea 
f_0(1-rf_0)\Big[1-\frac{(U\cdot p)}{h}\Big] p^{\mu}\nabla_{\mu}\alpha_{0}&=&\partial^{\beta}\alpha_{0} \bigg\{\bigg[K_{\parallel}\left( -\frac{U\cdot p}{\tau_{c}}\right)+\left(\frac{U\cdot p}{\tau_{c}}K_{\perp}+iqB K_{\times}\right)\bigg]b_{\nu}b_{\beta}p^{\nu}\nonumber\\  
 &-&\left[\frac{U\cdot p}{\tau_{c}}K_{\times}+iqB K_{\perp}\right]i b_{\nu\beta}p^{\nu} \Big\} -\left[\frac{U\cdot p}{\tau_{c}}K_{\perp}+iqB K_{\times}\right]p^{\mu}\nabla_{\mu}\alpha_{0}.
\eea  
\end{widetext}

Equating the coefficients for different tensorial terms we get,
\begin{align}
\left[\frac{U\cdot p}{\tau_{c}}K_{\perp}+iqB K_{\times}\right]&=f_0(1-rf_0)\Big[1-\frac{(U\cdot p)}{h}\Big],\nonumber \\
-\frac{U\cdot p}{\tau_{c}}K_{\parallel}+\frac{U\cdot p}{\tau_{c}}K_{\perp}+iqB K_{\times}&= 0,\nonumber\\
\frac{U\cdot p}{\tau_{c}}K_{\times}+iqB K_{\perp}&=0.
\end{align}

Equating the above three equations we get,
\bea 
K_{\parallel}=-\frac{\tau_{C}f_0(1-rf_0)}{U\cdot p}\Big[1-\frac{(U\cdot p)}{h}\Big] , \\
K_{\perp}=-\frac{\tau_{C}(U\cdot p)f_0(1-rf_0)}{(U\cdot p)^{2}+(qB \tau_{C})^{2}}\Big[1-\frac{(U\cdot p)}{h}\Big] ,  \\
K_{\times}=-\frac{qB \tau_{C}^{2}f_0(1-rf_0)}{(U\cdot p)^{2}+(qB \tau_{C})^{2}}\Big[1-\frac{(U\cdot p)}{h}\Big].
\eea 
So, the thermal diffusion coefficients $\kappa$'s become
\bea
\kappa_{\parallel}&=&-\frac{1}{3}\int  \frac{d^3{\vec p}}{(2\pi)^{3}p_{0}}{|\vp|}^{2}K_{\parallel},
\nn\\
&=&\frac{1}{3h}\int  \frac{d^3{\vec p}}{(2\pi)^{3}}\frac{|\vp|^2}{p_0^2}\tau_c(h-p_{0})f_0(1-rf_0)
%\tau_{c}\bigg(-\frac{n^{2}}{\beta(\epsilon+P)}+\nonumber\\
%&&\frac{2g}{3\beta}\int \frac{d^3{\vec p}}{{(2\pi)}^3 E} \left(1+\frac{m^2}{2E^2} (f_0-\bar{f_0})\right)\bigg)
\nn\\
\kappa_{\perp}&=&-\frac{1}{3}\int  \frac{d^3{\vec p}}{(2\pi)^{3}p_{0}}{|\vp|}^{2}K_{\perp},
\nn\\
&=&\frac{1}{3h}\int  \frac{d^3{\vec p}}{(2\pi)^{3}}\frac{|\vp|^2}{p_0^2}\frac{\tau_c(h-p_{0})}{1+(\frac{\tau_{c}}{\tau_{B}})^2}f_0(1-rf_0)
\nn\\
\kappa_{\times}&=&-\frac{1}{3}\int  \frac{d^3{\vec p}}{(2\pi)^{3}p_{0}}{|\vp|}^{2}K_{\times}
\nn\\
&=&\frac{1}{3h}\int  \frac{d^3{\vec p}}{(2\pi)^{3}}\frac{|\vp|^2}{p_0^2}\frac{\tau_c(\frac{\tau_{c}}{\tau_{B}})(h-p_{0})}{1+(\frac{\tau_{c}}{\tau_{B}})^2}f_0(1-rf_0)~.
\nn\\
\eea
%where
%{\color {magenta} Is this needed?
%\bea
%K_{\perp}& =& \frac{1}{1+(\tau_c/\tau_B)^{2}} K_{\parallel} \\
%K_{\times} &=&  \frac{(\tau_c/\tau_B)}{1+(\tau_c/\tau_B)^{2}} K_{\parallel}~.
%\eea
%}

\end{document}